# Phylogenomic Analyses of Large-scale Nuclear Genes Provide New Insights into the Evolutionary Relationships within the Rosids

Lei Zhao[a, b], Xia Li[a], Ning Zhang[c], Shu-Dong Zhang[a], Ting-Shuang Yi[a], Hong Ma[d], Zhen-Hua Guo [a, *], De-Zhu Li[a, *]

[a]*Plant Germplasm and Genomics Center, Germplasm Bank of Wild Species, Kunming Institute of Botany, Chinese Academy of Sciences, Kunming, Yunnan 650201, China*

[b]*Kunming College of Life Sciences, University of Chinese Academy of Sciences, Kunming, Yunnan 650201, China*

[c]*Department of Botany, National Museum of Natural History, MRC 166, Smithsonian Institution, Washington, DC 20013-7012, USA*

[d]*Ministry of Education Key Laboratory of Biodiversity Sciences and Ecological Engineering, Institute of Plant Biology, Institute of Biodiversity Sciences, Center for Evolutionary Biology, School of Life Sciences, Fudan University, Shanghai 200032, China*

*Correspondence: De-Zhu Li, Kunming Institute of Botany, Chinese Academy of Sciences, 132 Lanhei Road, Kunming, Yunnan 650201, China. Tel: +86 871 6522 3503.

E-mail: dzl@mail.kib.ac.cn

Zhen-Hua Guo, Kunming Institute of Botany, Chinese Academy of Sciences, 132 Lanhei Road, Kunming, Yunnan 650201, China. Tel: +86 871 6522 3153.

E-mail: guozhenhua@mail.kib.ac.cn

# Abstract

The Rosids is one of the largest groups of flowering plants, with 140 families and ~70,000 species. Previous phylogenetic studies of the rosids have primarily utilized organelle genes that likely differ in evolutionary histories from nuclear genes. To better understand the evolutionary history of rosids, it is necessary to investigate their phylogenetic relationships using nuclear genes. Here, we employed large-scale phylogenomic datasets composed of nuclear genes, including 891 clusters of putative orthologous genes. Combined with comprehensive taxon sampling covering 63 species representing 14 out of the 17 orders, we reconstructed the rosids phylogeny with coalescence and concatenation methods, yielding similar tree topologies from all datasets. However, these topologies did not agree on the placement of Zygophyllales. Through comprehensive analyses, we found that missing data and gene tree heterogeneity were potential factors that may mislead concatenation methods, in particular, large amounts of missing data under high gene tree heterogeneity. Our results provided new insights into the deep phylogenetic relationships of the rosids, and demonstrated that coalescence methods may effectively resolve the phylogenetic relationships of the rosids with missing data under high gene tree heterogeneity.

**Keywords**:

Coalescence; Rosids phylogenomics; Nuclear genes; Gene tree; Species tree

# 1. Introduction

The rosids is one of the most diverse lineages of flowering plants, containing 17 orders, which in turn comprise 140 families and ca. 70,000 species, exhibiting remarkable morphological and ecological diversities (APG II, 2003; APG III, 2009; Chase et al., 1993; Magallon et al., 1999). The rosids consists of an unusually heterogeneous group with respect to habitat and life form, with member species occurring as herbs, trees, aquatics and succulents. Some members are significant cash crops (e.g., Fabaceae, Rosaceae and Brassicaceae), and others are important forest trees (e.g., Betulaceae, Fagaceae and Sapindaceae). In previous studies based on chloroplast and mitochondrial genes, the rosids has been divided into two major clades (Fig. 1a): (i) the fabids, which contains the nitrogen-fixing clade (Cucurbitales, Fagales, Fabales and Rosales), the COM clade (Celastrales, Oxalidales and Malpighiales) and Zygophyllales; and (ii) the malvids, which includes Brassicales, Malvales, Sapindales, Crossosomatales, Picramniales, Huerteales, Geraniales and Myrtales (APG IV, 2016; Hilu et al., 2003; Judd and Olmstead, 2004; Moore et al., 2011; Qiu et al., 2010; Soltis et al., 2011; Wang et al., 2009; Zhu et al., 2007).

Despite these prior works, the positions of some clades, the COM clade, Geraniales, Myrtales and Zygophyllales remained uncertain (APG IV, 2016; Maia et al., 2014; Morton, 2011; Sun et al., 2015). The COM clade, as circumscribed by these two studies (Endress and Matthews, 2006; Zhu et al., 2007), was sister to the nitrogen-fixing clade of the fabids according to some previous studies based on chloroplast genes (Burleigh et al., 2009; Hilu et al., 2003; Jansen et al., 2007; Moore et al., 2010; Ruhfel et al., 2014; Soltis et al., 2007; Soltis et al., 2011; Soltis et al., 2000; Wang et al., 2009). Subsequently, relying on mitochondrial (Qiu et al., 2010; Zhu et al., 2007) and nuclear genes (Burleigh et al., 2011; Duarte et al., 2010; Lee et al., 2011; Maia et al., 2014; Xi et al., 2014; Zhang et al., 2012), other studies placed the COM clade as a part of the malvids. In addition, while the COM clade formed a monophyletic group, within it M-O (Qiu et al., 2010; Ruhfel et al., 2014; Soltis et al., 2011; Soltis et al., 2000; Wang et al., 2009; Wu et al., 2014; Zhu et al., 2007), M-C (Burleigh et al., 2009; Moore et al., 2011; Zhang and Simmons, 2006) and O-C (Hilu et al., 2003; Moore et al., 2010; Ruhfel et al., 2014; Sun et al., 2016) were respectively supported as sister groups by different studies (Sun et al., 2015). With respect to the placements of Geraniales and Myrtales, in some studies using mitochondrial genes, they were supported as successive sisters to the remaining rosids (APG II, 2003; Hilu et al., 2003; Qiu et al., 2010; Zhu et al., 2007). However, in other studies based on chloroplast genes, they were placed in the malvids (APG IV, 2016; Jansen et al., 2007; Ruhfel et al., 2014; Soltis et al., 2011; Wang et al., 2009). For the position of Zygophyllales, the group has been placed in the malvids (Maia et al., 2014; Qiu et al., 2010; Ruhfel et al., 2014) or the fabids (Hilu et al., 2003; Ruhfel et al., 2014; Soltis et al., 2011; Wang et al., 2009). Recently, a few studies have investigated the positions of some uncertain orders by using nuclear genes, albeit with limited taxon sampling. They provided supports for grouping of some COM orders with malvids (Finet et al., 2010; Lee et al., 2011; Sun et al., 2015; Xi et al., 2014; Zhang et al., 2012) and placing Myrtales (or together with Geraniales) as sister to the remaining rosids (Myburg et al., 2014; Sun et al., 2015; Wang et al., 2014; Zeng et al., 2014).

Currently, due to increased affordability, high-throughput sequencing technologies have been widely employed for genome and transcriptome sequencing (Reuter et al., 2015). They allow data on hundreds or thousands of single or low copy nuclear genes to be collected for inferring species relationships (Lemmon and Lemmon, 2013; Wen et al., 2015; Zimmer and Wen, 2015). However, with such large, genome-scale datasets, phylogenetic conflicts among genes become evident, resulting in phenomena such as gene tree-species tree discordance (Degnan and Rosenberg, 2009; Salichos and Rokas, 2013; Szollosi et al., 2015). Therefore, simply increasing the number of gene sequences does not always resolve phylogenetic incongruences (Kimball and Braun, 2014; Nater et al., 2015; Philippe et al., 2011). Additionally, revisitations of previously published phylogenomic datasets using different analytical methods often produce conflicting results (Simmons and Gatesy, 2015; Springer and Gatesy, 2014, 2016; Tarver et al., 2016), indicating that the choice of analytical methods is an important consideration for the phylogenetic studies (Roch and Warnow, 2015).

In traditional phylogenetic analyses, multiple genes are concatenated as a supermatrix for inferring evolutionary relationships (de Queiroz and Gatesy, 2007). This concatenation method has been widely employed in numerous phylogenomic studies of animals (Jarvis et al., 2014; Kocot et al., 2011), plants (Wickett et al., 2014; Zeng et al., 2014) and fungi (Ebersberger et al., 2012; Spatafora and Bushley, 2015). Concatenation methods assume that all genes have the same evolutionary history, ignoring or downplaying inevitable evolutionary processes, such as incomplete lineage sorting (ILS), horizontal gene transfer (HGT), and gene duplication and loss (GDL) (Degnan and Rosenberg, 2009; Knowles, 2009; Nakhleh, 2013). Species tree estimation from large multi-locus datasets could be complicated by these biological processes (Edwards, 2009; Kutschera et al., 2014; Lambert et al., 2015; Som, 2015), because they cause gene tree heterogeneity, which are not explicitly accounted for by concatenation methods (Knowles, 2009; Nosenko et al., 2013; Salichos and Rokas, 2013; Szollosi et al., 2015). Other issues that may complicate phylogenetic estimation are substitution saturation, long-branch attraction (LBA) and missing data, although these issues are not necessarily exclusive to concatenation methods (Liu et al., 2015b; Roure et al., 2013; Whelan et al., 2015; Xi et al., 2016). As a result of these issues, concatenation methods may introduce significant errors or produce highly supported but incorrect species tree topologies (Giarla and Esselstyn, 2015; Linkem et al., 2016; Roch and Steel, 2014; Zhong et al., 2013).

Recently, many coalescence methods have been developed to address these problems (Knowles, 2009; Liu et al., 2015a; Szollosi et al., 2015). The first type of methods is termed co-estimation methods, e.g., BEST (Liu, 2008) and *BEAST (Heled and Drummond, 2010), which simultaneously infer gene trees and the underlying species tree. These methods have outstanding accuracy, but are computationally demanding for large datasets (Leache and Rannala, 2011; Mirarab et al., 2014b). The second type of methods is called single-site methods and they use single nucleotide polymorphisms (SNPs) to infer species trees. Examples of software that implement this type of methods are SNAPP (Bryant et al., 2012) and SVDquartets (Chou et al., 2015). The third type of methods is called summary methods. Relying on the multi-species coalescent model, methods in this class produce statistically consistent estimation of the true species tree using multiple gene trees as input (Liu et al., 2009a; Mirarab et al., 2014b). These methods converge on true species tree with increasing amounts of data (Knowles and Kubatko, 2011; Liu et al., 2009a; Ruane et al., 2015). Software that implement summary methods include ASTRAL (Mirarab et al., 2014c), ASTRAL-II (Mirarab and Warnow, 2015), STAR (Liu et al., 2009b), MP-EST (Liu et al., 2010), NJst (Liu and Yu, 2011), STEM (Kubatko et al., 2009) and ASTRID (Vachaspati and Warnow, 2015). The fourth type of methods is called statistical binning methods, which are hybrid approaches in which subsets of loci are grouped into bins based on a statistical test before summary coalescent methods are applied. One software that uses this method is *MP-EST (Bayzid et al., 2015; Mirarab et al., 2014a). Recent simulated and empirical studies have shown that coalescence methods are superior to traditional concatenation methods under high ILS (Davidson et al., 2015; Linkem et al., 2016; Liu et al., 2015b; Mirarab et al., 2014b; Xi et al., 2016). However, there is still an ongoing debate regarding which approach is more appropriate to construct species trees when using large-scale phylogenomic datasets (Edwards et al., 2016; Springer and Gatesy, 2016; Tonini et al., 2015; Warnow, 2015).

In traditional studies of the rosids, chloroplast and mitochondrial genetic markers are mainly used in conjunction with traditional phylogenetic analyses. Generally, chloroplast and mitochondrial genes in plants are uniparentally inherited but are affected by rampant HGT, which might lead to biases when inferring phylogenetic relationships (Birky, 2001; Davis et al., 2014). In contrast, nuclear genes are inherited biparentally, which could supply alternative evidence for the rosids phylogeny. In this study, we generated a large-scale phylogenomic data composed of 891 orthologous gene clusters. Sixty three species representing 14 out of the 17 orders within the rosids were sampled (Fig. 1a and Table S1). The phylogenetic relationships among major orders were reconstructed by concatenation and coalescence methods. In this study, our main goals were to: (1) reinvestigate the deep phylogenetic relationships among some major orders using large-scale nuclear gene data; and (2) determine the position of some orders with previously uncertain placements.

## 2. Materials and Methods

### 2.1 Taxon Sampling

Our taxon sampling included 63 rosids species representing 14 out of the 17 orders, each of which contained at least two families, except for Celastrales, Geraniales and Zygophyllales. The sampled taxa were listed in Supplementary Table S1. Among these, we downloaded publicly available genomes (36 species) and RNA-Seq data (26 species) from Phytozome (http://phytozome.jgi.doe.gov), NCBI (http://www.ncbi.nlm.nih.gov/), and GigaDB (http://gigadb.org). We generated new Illumina paired-end RNA-Seq data of *Oxalis corymbosa* (Oxalidales, Oxalidaceae). Roots, stems, leaves and flowers tissue samples were obtained from three populations, and immediately frozen and stored in liquid nitrogen for RNA extraction. Total RNA was extracted from pooled materials using the Plant Total RNAs Extraction Kit 3301. To obtain pure RNA, residual DNA in the total RNA samples was digested by RNase-free DNase I (Takara, Japan). RNA-Seq library construction, Illumina HiSeq 2000 sequencing, raw data cleaning and quality control were performed at BGI Shenzhen, China.

## 2.2 Sequence assembly and ortholog identification

We employed an integrated bioinformatics pipeline designed for large-scale datasets, as shown in Supplementary Fig S1. Trinityrnaseq-2.0.2 and Newbler2.9 were used to conduct *de novo* assembly of Illumina or Roch 454 RNA-Seq reads of each species, respectively (Haas et al., 2013; Margulies et al., 2005). To obtain non-redundant transcript sequences, contigs of each species were further clustered using the TGI Clustering tool (Pertea et al., 2003). TransDecoder_r20140704 was subsequently employed to predict CDS region, and sequences ≤300bp were discarded (Haas et al., 2013). To identify orthologous clusters from genomes and transcriptomes, HaMStRv8 (http://www.deep-phylogeny.org/hamstr/) was performed with strict parameters (-hmmset=magnoliophyta_hmmer3, -representative, -strict, -eval_limit= 0.00001, and

-rbh). When conducting HaMStR analysis, we used core angiosperm orthologs (4,180 core orthologs) made up of data from 'primer taxa', which included *Arabidopsis thaliana*, *Glycine max*, *Medicago truncatula*, *Populus trichocarpa*, *Oryza sativa*, *Sorghum bicolor*, *Solanum lycopersicum*, *Vitis vinifera* and *Zea mays* (Ebersberger et al., 2009; Zhao et al., 2013). Our search of ortholog clusters resulted in 891 clusters from 63 species. To reduce missing data and to balance taxon sampling in each cluster of orthologous gene, we only included clusters that were represented by at least 60% of the species.

## 2.3 Matrix construction and species tree inference

Each cluster was aligned using TranslatorX with MAFFT, and poorly aligned regions were trimmed by trimAlv1.4 with default parameters (Abascal et al., 2010; Capella-Gutierrez et al., 2009). Following alignment, we carried out phylogenetic inferences with concatenation and summary coalescence methods.

All trimmed alignments were concatenated using SCaFoS (Roure et al., 2007). We concatenated the full dataset and a variety of sub-datasets to generate different supermatrices. First, Matrix A consisted of all codon positions (nt123) from 891 clusters of 63 taxa. To take into account rate heterogeneity, concatenation analyses employed the GTR+CAT model when maximum-likelihood (ML) inference was conducted. ML trees were constructed using RAxML-8.1 under the following settings: 500 bootstrap replicates, GTRCAT and "-f a" option (Stamatakis, 2006).

It is known that concatenation methods can be statistically inconsistent or even positively misguiding under high ILS (Kubatko and Degnan, 2007; Roch and Steel, 2014). Thus, we further constructed a species tree by coalescence methods based on best-scoring ML gene tree of each constituent orthogroup of the supermatrix dataset. Each gene tree was inferred using RAxML8.1 with GTRGAMMA model, 100 bootstrap replicates and a rapid bootstrapping algorithm (-f option). The species tree was then constructed from the best-scoring ML gene trees using ASTRAL4.7.7 (Mirarab and Warnow, 2015). Multi-locus bootstrapping analyses were implemented for best-scoring ML gene trees and ML bootstrap replicates (-b and -r option).

## 2.4 The assessment of incongruent species tree

Generally, the third codon position evolves relatively rapidly and easily reaches saturation, thus influencing phylogenetic estimation. To assess whether the rapid evolutionary rate of the third codon position resulted in the conflicting positions of Zygophyllales (see details in the Result section), we constructed Matrix B based on only the first and second codon positions (nt12). Additionally, we identified fast-evolving and slow-evolving sites of each gene using TIGER-v1.02 (Cummins and McInerney, 2011). These sites were then concatenated to Matrix C and D, respectively.

LBA is a well-known phenomenon of spurious species tree inference with long branches grouping together (Felsenstein, 1978). It often occurs when there is sparse taxon sampling and a presence of substitution saturation (Egger et al., 2015; Liu et al., 2015b). To investigate whether LBA resulted in the incongruent placements of Zygophyllales, we removed representative taxa of Geraniales and Myrtales from Matrix A to generate Matrix E.

Missing data arise when there are various factors in phylogenomic datasets: 1) orthologous genes having been lost or not being identified; 2) insertions, deletions, and other chromosomal variations being included in the genomes; and 3) variable sequences among species yielding missing data across multiple sequence alignments (Lemmon et al., 2009; Xi et al., 2016). The impacts of missing data on phylogenetic incongruence have been detected in large-scale datasets by some previous studies (Dell'Ampio et al., 2014; Kvist and Siddall, 2013; Roure et al., 2013). To investigate the effects of missing data on the phylogenetic position of Zygophyllales, 604 orthologous clusters out of a total of 891 were extracted. Data from *Larrea tridentata* (a member of Zygophyllales) were represented in each of these clusters. In other words, 287 other orthologous genes may be lost or cannot be detected from the transcriptomic datasets of *L. tridentata*. Subsequently, we analyzed subsets of this dataset created using the following criteria: 1) less than 10% data missing (303 clusters), 2) less than 5% data missing (273 clusters), 3) less than 1% data missing (213 clusters) from the initial dataset containing the 604 orthologous clusters.

Additionally, gene tree heterogeneity is an important source of phylogenetic conflicts in multi-locus nuclear datasets (Liu et al., 2015c). To evaluate the level of gene tree heterogeneity in our datasets, we estimated pairwise Robinson-Foulds (RF) distances (Robinson and Foulds, 1981; Simmons et al., 2016) among gene trees from the datasets that contained 891, 604, 303, 273 and 213 clusters. The RF distance shows what percentage of clades is different between two gene trees.

Finally, we examined how anomalous gene trees impacted phylogenetic inferences. Based on widely supported phylogenetic relationships derived from studies which employed plastid genes (Soltis et al., 2011; Wang et al., 2009), mitochondrial genes (Qiu et al., 2010; Zhu et al., 2007), nuclear genes (Xi et al., 2014; Zhang et al., 2012) and floral structural characters (Endress, 2010; Endress and Matthews, 2006), we removed “problematic” gene trees using on the following rules: (1) Representative taxa of Fabales, Fagales, Rosales and Cucurbitales cannot be nested within Malvales, Brassicales and Sapindales, and *vice versa* (Fig. 1b);

(2) Species that belong to the same family should form a clade, and

(3) Species that belong to the same genus should form a clade.

# 3 Results

## 3.1 Overwhelmingly consistent topologies among species trees

Matrix A was composed of 1,120,686 nucleotide sites with 22.37% missing data. Based on Matrix A, ML concatenation analysis produced a highly supported phylogenomic tree, and most nodes received 100% bootstrap support (BS) values (Fig. 2). The fabids included only the nitrogen-fixing clade, within which Fabales and Fagales grouped together, and so did Rosales and Cucurbitales. The orders of the COM clade did not form a monophyletical group. Of the three COM orders, Celastrales was sister to Malpighiales, while Oxalidales was placed as sister to the malvids. Within the malvids, Malvales was sister to Brassicales rather than Sapindales. Geraniales was sister to the Myrtales-Zygophyllales clade, and the clade formed by these three orders was in turn sister to the remaining rosids. The species tree derived from coalescence analysis of gene trees of Matrix A was completely congruent with that produced by concatenation analysis, except for the phylogenetic position of Zygophyllales (Fig. 2). Under coalescence analysis, Zygophyllales was strongly supported as the basal lineage of the “expanded” malvids (including Zygophyllales, Celastrales, Malpighiales, Oxalidales, plus other “traditional” malvid orders) with 100% bootstrap support.

## 3.2 The causes of phylogenetic incongruence

To address the causes of the discordant position for Zygophyllales, we further carried out in-depth analyses of some factors, such as presence of fast-evolving sites, LBA, level of missing data and gene tree heterogeneity.

Using ML concatenation analysis, we inferred phylogenetic tree from Matrix B (761,630 sites matrix, 22.58% missing data). The tree topology we obtained was the same as that generated from Matrix A by concatenation method (Fig. S2). Subsequently, the species tree was constructed from constituent orthogroups of Matrix B by coalescence method. It matched the tree generated by concatenation method, except that Zygophyllales and the Geraniales-Myrtales clade were sister to the “expanded” malvids with weak support (Fig. S3). We also parsed fast and slow sites from each of the 891 orthologous gene clusters using TIGER-v1.02. Regardless, the position of Zygophyllales was still different between concatenation and coalescence methods (Figs. S4-S7). To avoid the possible impact of LBA, ML tree was also inferred from Matrix E (1,120,686 sites matrix, 22.24% missing data). However, this did not change the placement of Zygophyllales, and it remained sister to the remaining rosids (Fig. S8).

We calculated summary statistics of missing data from the five *L. tridentata* supermatrices, as shown in Table 1. Zygophyllales was strongly supported as sister to Myrtales based on concatenation analyses of 891 and 604 clusters, and BS values of the relationship were 90% and 95%, respectively (Table 1, Figs. 2 and S9). However, the two supermatrices included large amounts of missing data (52% and 28%). Subsequently, starting with the set of 604 clusters whereby data of *L. tridentata* were represented in each cluster, we reduced the total rate of missing data of each cluster to less than 10%, less than 5% and less than 1%. This reduced the total number of clusters to 303, 273 and 213, respectively. In the two supermatrices of 303 and 273 clusters, the amount of missing data decreased substantially with overall rates of only 2% and 0.7%, respectively. Meanwhile, the support for the Zygophyllales-Myrtales clade dropped from 59% to 20% (Table 1, Figs. S10, S11). In the supermatrix of 213 clusters, the overall missing data rate was only 0.2%. Based on concatenation analysis of this supermatrix, the phylogenetic position of Zygophyllales changed. Zygophyllales was now basal to the "expanded" malvids (albeit with only 45% bootstrap support), which was congruent with the results of all coalescence analyses (Table 1, Fig. S12). In the process of conducting coalescence analyses, the position of Zygophyllales was very stable and remained be basal to the "expanded" malvids with strong support (Table 1, Figs. 2 and S13-S16). These findings indicate that high missing data rate may have significant impacts on phylogenetic inferences that use large phylogenomic datasets, and may indeed misguide concatenation methods (Roure et al., 2013; Simmons, 2012, 2014).

To further address sources of the observed phylogenetic conflicts, we also analyzed gene tree heterogeneity by calculating RF distances among gene trees within each of the five datasets (Fig. 3). The distributions of RF distances were mostly between 0.35 and 0.55. None of five datasets contained pairs of gene trees with RF distance of <0.3. That is, clades in each gene tree of each dataset were in conflict at least 30% of the time when the gene tree was compared against all other trees of the same dataset. From Fig. 3, we also observed gene tree heterogeneity decreasing gradually when the numbers of clusters were reduced from 891 to 213. The combination of Fig. 3 and Table 1 showed that under concatenation method, the position stability of Zygophyllales fell as gene tree heterogeneity was lowered and missing data reduced, which was in striking contrast to coalescent method. These results indicated that coalescence method could reconcile gene tree heterogeneity, and be robust to missing data in orthologous genes (Liu et al., 2015a; Roch and Warnow, 2015; Tonini et al., 2015; Warnow, 2015). In contrast, concatenation method was misguided by the combination of high missing data rate and high gene tree heterogeneity, even while bootstrap support for the Zygophyllales-Myrtales sister relationship remained high (90% and 95% BS from 891 and 604 clusters, respectively) (Xi et al., 2016).

Finally, through examination of individual gene trees, we found a large number of anomalous gene trees with "problematic" topologies, which might have been affected by LBA, ILS or undetected paralogy (Figs. S17-S19). According to our rules (Materials and Methods), we excluded a large number of gene tress with "problematic" topologies, and retained a final set of 193 gene trees from an initial set of 891. Subsequent to gene tree filtering, pairwise RF distances of the 193 gene trees were also computed. The distributions of RF distances were only from 0.2 to 0.35. Compared with previous datasets, gene trees heterogeneity decreased substantially (Figs. 3, 4). Interestingly, concatenation and coalescence analyses of this final set yielded identical tree topologies, which in turn were congruent with the phylogeny produced by coalescence analyses of the total dataset and by concatenation analysis of the dataset containing 213 clusters (for each gene found in *L. tridentata*, there was less than 1% missing data), albeit with relatively low support from concatenation methods (Figs. 2, 5 and S12). The results again suggest that coalescence method is more robust than concatenation method when inferring species tree when high gene tree heterogeneity is a problem.

# 4 Discussion

## 4.1 The phylogenetic incongruence, concatenation and coalescence methods

There are substantial controversies over whether to employ concatenation methods or coalescence methods due to their respective advantages and disadvantages. Some analyses based on simulated or empirical data showed that concatenation methods provided better accuracies than coalescence methods under low enough ILS, weak phylogenetic signal and a small number of genes (Gatesy and Springer, 2014; Patel et al., 2013; Springer and Gatesy, 2016; Warnow, 2015). However, concatenation methods can be misguided by a considerable amount of missing data (Kvist and Siddall, 2013; Roure et al., 2013; Simmons, 2014), LBA (anomaly zone) (He et al., 2016; Linkem et al., 2016; Whelan et al., 2015), substitution saturation (fast evolution sites) (Chiari et al., 2012; Lin et al., 2014) and high ILS (Liu et al., 2015b; Warnow, 2015; Xi et al., 2016). Hence, concatenation methods, which assume that all genes are homogeneous, are not statistically inconsistent under these circumstances (Kubatko and Degnan, 2007; Roch and Steel, 2014; Warnow, 2015). Other studies demonstrated that summary coalescent methods outperformed concatenation methods in some conditions, especially under substantial missing data and high ILS (Davidson et al., 2015; Liu et al., 2015b; Ruane et al., 2015; Tonini et al., 2015; Xi et al., 2016). Generally, coalescence methods accommodate gene tree heterogeneity, and the performance can be improved by increasing the quantity of gene trees (Camargo et al., 2012; Knowles and Kubatko, 2011; Roch and Warnow, 2015). Nevertheless, summary coalescent methods may result in less accurate species trees, when gene trees have estimation uncertainties due to short, uninformative and low quality genes (Lanier et al., 2014; Patel et al., 2013; Szollosi et al., 2015; Xi et al., 2015).

In our study, concatenation and coalescence analyses of the total dataset composed of 891 clusters yielded mostly consistent topologies, but conflicting results for the position of Zygophyllales, which was a sister of Myrtales in concatenation analyses but was basal to the “expanded” malvids in coalescence analyses. One potential factor behind this incongruence is that the supermatrix contains a large amount of missing data. Although next-generation sequencing platform produced large amounts of sequencing data, concatenation of multiple genes into a single supermatrix may still result in a high missing data rate due to unrecognized orthologs, differential gene expression, and insertions and deletions (Kvist and Siddall, 2013; Yeates et al., 2016). In this study, we analyzed the impact of missing data, and constructed species tree from five supermatrices resulting from the concatenation of 891 clusters, 604 clusters and three clusters with less than 10% (303 clusters), 5% (273 clusters), and 1% (213 clusters) missing data in each gene of *L. tridentata*, respectively. The results showed that the placement stability of Zygophyllales was significantly reduced as the amount of missing data decreased (Table 1, Figs. 2, S9-S12). When the supermatrix employed to infer the phylogenetic tree contained less than 1% missing data in *L. tridentata* gene, the position of Zygophyllales changed from being sister of Myrtales to being basal of the “expanded” malvids, which was in accordance with that produced by coalescence methods based on all five supermatrices. This finding suggested that missing data may have potential impacts on the position of Zygophyllales for concatenation method (Simmons, 2014; Xi et al., 2016). Although some studies also suggested that coalescence methods were remarkably resilient to the issue of missing data, the amount of missing data will have a negligible impact on species tree estimation as the number of gene increased (Hovmoller et al., 2013; Streicher et al., 2015). Another potential confounding factor is gene tree heterogeneity. Generally, concatenation methods assume that all genes evolve along the same or similar evolutionary histories, and overlook gene tree heterogeneity. To measure gene tree heterogeneity, we looked at pairwise RF distances within each of five datasets, and found that gene tree heterogeneity was very high for each dataset. Under concatenation method, the positional stability of Zygophyllales changed when gene tree heterogeneity lessened (Table 1, Figs. 3, 4 and S9-S12). However, under coalescence analysis, the placement of Zygophyllales as the basal lineage of the "expanded" malvids remained stable regardless of the dataset used (Table 1, Figs. 3, 4 and S13-S16). These results suggested that gene tree heterogeneity had potential influences on the position of Zygophyllales under concatenation method. Above all, we illustrated that concatenation methods may produce highly misleading results when faced with large-scale missing data under high gene tree heterogeneity. However, summary coalescence methods also have bias. In the presence of substantial gene tree error, coalescence methods may give rise to inaccurately inferred species tree as well (Meiklejohn et al., 2016; Xi et al., 2015). We manually inspected each gene tree from the 891 clusters, and confirmed that there was a large set of heterogeneous gene trees with different topologies. After erroneous gene trees were excluded, analyses of the remaining 193 clusters further demonstrated that coalescence method was more efficient than concatenation methods for inferring species tree when high gene tree heterogeneity was present in our phylogenomic datasets (Figs. 3-5).

In summary, hundreds of nuclear genes provided us with an opportunity to explore incongruence between gene trees and species tree, as well as addressed the possible sources of conflicts for analytical methods. Our work indicates that concatenation and coalescence methods should both be employed for the phylogenetic analysis of large nuclear DNA datasets in the future. Furthermore, when there are conflicts in results derived from concatenation and coalescence methods, we should consider some potential factors like the level of missing data and gene tree heterogeneity (Posada, 2016).

## 4.2 The phylogenetic relationships of rosids

In most studies using organelle genes, the rosids consists of two major clades: the fabids and the malvids. We also recovered two similar clades, and indicated that these phylogenetic relationships were the most strongly supported by organelle and nuclear genes. In our results, however, the fabids only contained the nitrogen-fixing clade. The Myrtales-Geraniales clade was sister to the remaining rosids. Zygophyllales, the CM clade, Oxalidales and other malvids orders composed the "expanded" malvids.

### 4.2.1 The Malpighiales-Celastrales clade and Oxalidales

Previous studies based on plastid genes suggested that the COM clade was nested within the fabids (Fig. 1a) (Burleigh et al., 2009; Hilu et al., 2003; Moore et al., 2010; Ruhfel et al., 2014; Soltis et al., 2007; Soltis et al., 2011; Wang et al., 2009; Wu et al., 2014). However, previous analyses based on mitochondrial genes supported closer relationships between the malvids and the COM clade (Qiu et al., 2010; Zhu et al., 2007). Recently, some studies investigated placement of the COM clade by using nuclear genes, but sampling of these three orders were not complete (Lee et al., 2011; Shulaev et al., 2011; Xi et al., 2014; Zeng et al., 2014; Zhang et al., 2012). In this study, our taxon sampling of the three orders was complete. In our results, the COM orders did not form a monophyletic group. A sister relationship of M-C was strongly supported but not M-O and O-C, finding that has been reported by Zhang and Simmons (2006) based on analyses of plastid and nuclear genes. The M-C clade and Oxalidales were members of the malvids rather than the fabids (Figs. 2 and 5), which were consistent with studies based on nuclear (Morton, 2011; Wickett et al., 2014; Zeng et al., 2014) and mitochondrial genes (Qiu et al., 2010; Zhu et al., 2007). Additionally, the sister relationship between Malpighiales-Celastrales, Oxalidales and the malvids were also well-supported by a wide-ranging survey of floral structural characters e.g., ovule, contort petals (Endress, 2010; Endress and Matthews, 2006; Schonenberger and von Balthazar, 2006). Nevertheless, incongruence in the positions of the COM orders between studies based on chloroplast, mitochondrial and nuclear genes remained a puzzle. Sun et al. (2015) indicated that ancient hybridization and introgression event could be plausible explanations.

#### 4.2.2 Zygophyllales

Zygophyllales as a taxon was first recognized and adopted in these studies (APG III, 2009; Soltis et al., 2000). Most previous studies based on plastid genes supported its sister relationship to the rest of the fabids (the COM and N-fixing clades) (APG II, 2003; APG III, 2009; APG IV, 2016; Moore et al., 2011; Moore et al., 2010; Soltis et al., 2011; Sun et al., 2015; Wang et al., 2009; Wu et al., 2014). In spite of this, its position was very variable. Hilu et al. (2003) using *matK* sequences suggested that Zygophyllales was sister to Fabales with weak support. A study by Burleigh et al. (2009), which used 5 genes (18S rDNA, *atpB*, *rbcl*, *matK*, and 26S rDNA), provided weak evidence for the relationships between Zygophyllales and the COM clade. Analyses of (Ruhfel et al., 2014) 78 plastid genes appeared to support a rather uncertain position of Zygophyllales. Additionally, some studies indicated that it belonged to the malvids. Wang et al. (2009), using a maximum parsimony method, suggested that Zygophyllales was sister to the malvids, although an AU test rejected the position of Zygophyllales. Zhu et al. (2007), using mitochondrial *matR* gene, indicated that Zygophyllales was at the base of the malvids (including the COM clade). Qiu et al. (2010), based on four mitochondrial genes (*atp1*, *matR*, *nad5* and *rps3*), suggested that Zygophyllales was embedded in Crossosomatales, the combined clade being sister to the remaining of rosids. In our study, the placement of Zygophyllales was not consistent between coalescence and concatenation methods. Through deep analysis, our results strongly supported that Zygophyllales was at the base of the "expanded" malvids (Table 1, Figs. 2, 5 and S13-S16). This position was consistent with the findings of Wickett et al. (2014) based on 852 nuclear genes and coalescence analyses, and Zhu et al. (2007) based on mitochondrial *matR* gene. Nevertheless, our sampling only included *L. tridentata* in Zygophyllales. In further, more taxa should be sampled.

#### 4.2.3 The Nitrogen-fixing clade

Within the nitrogen-fixing clade, the phylogenetic relationships among the four clearly monophyletic orders remained unclear in previous studies. Although some studies based on chloroplast genes supported a sister relationship of Cucurbitales-Fagales (Burleigh et al., 2009; Moore et al., 2011; Moore et al., 2010; Soltis et al., 2011; Wang et al., 2009), this was poorly supported in other studies using mitochondrial genes or floral structural characters (Endress and Matthews, 2006; Qiu et al., 2010; Zhu et al., 2007). However, based on nuclear genes, the phylogenetic relationships of Fabales-Fagales (Zhang et al., 2012), Rosales-Cucurbitales (Xi et al., 2014) were well supported, respectively. In our results, the sister relationships of Fabales-Fagales, and Rosales-Cucurbitales were strongly supported with 100% bootstrap value from both concatenation and coalescence methods (Figs. 2 and 5). These relationships were in agreement with those of Nickrent et al. (2005) based on plastid *rbcL*, *atpB*, mitochondrial *matR* and nuclear SSU rDNA sequences.

### 4.2.4 Geraniales and Myrtales

The positions of Geraniales and Myrtales were variably placed by previous analyses. Therefore, the placements of Geraniales and Myrtales were not well resolved (Myburg et al., 2014; Soltis et al., 2007; Sun et al., 2015; Wang et al., 2009; Zhu et al., 2007). Using chloroplast genes, Wang et al. (2009) proposed that they belonged to the malvids (Jansen et al., 2007; Moore et al., 2011; Ruhfel et al., 2014; Soltis et al., 2011). Based on four mitochondrial and chloroplast genes, Myrtales and Geraniales were supported as successively sister to all other rosids, albeit with weak support (Qiu et al., 2010; Zhu et al., 2007). Recently, a few studies have placed Myrtales as sister to the remaining rosids, although representative taxa of Geraniales were not included and species sampling was limited (Morton, 2011; Myburg et al., 2014; Wang et al., 2014; Xi et al., 2014; Zeng et al., 2014). Our study included representative taxa of both orders. Geraniales and Myrtales formed a sister group, which in turn formed a strongly supported basal lineage to the remaining rosids (Figs. 2, 5). The result was consistent with those of Zhang et al. (2012) using five nuclear genes.

## 5 Conclusion

In this study, by using coalescence and concatenation analyses, our results provided new insights into the rosids relationships: 1) A clade made up of Myrtales-Geraniales was sister to the remaining orders within the rosids; 2) The fabids only included the nitrogen-fixing clade (Fabales-Fagales, Rosales-Cucurbitales); 3) The COM clade was not monophyletic, Celastrales and Malpighiales were sister to each other, and the clade was sister to Oxalidales plus the malvids; and 4) Zygophyllales was the basal to “expanded” malvids rather than to the fabids. Overall, our phylogenomic analyses provided valuable insights into the phylogenetic relationship of rosids, and suggested that coalescence methods, e.g., ASTRAL-II, were more efficient than concatenation methods in coping with large amounts of missing data under high gene tree heterogeneity.

## Acknowledgments

We thank Peng-Fei Ma, Yu-Xiao Zhang, Jun He, Guo-Qian Yang, and Hui-Fu Zhuang of Kunming Institute of Botany, Chinese Academy of Sciences for help and computational supports. We are grateful to Bo-Jian Zhong of Nanjing Normal University for his suggestions. This study is funded by the National Key Basic Research Program (No. 2014CB954100) and Kunming Institute of Botany, Chinese Academy of Sciences (No. 2014KIB02).

## Appendix A. Supplementary material

Supplementary data associated with this article can be found, in the online version, at http://dx.doi.org/xxx.

# Figure legends

**Fig. 1.** The phylogeny of 17 orders within the rosids is redrawn from Wang et al. (2009), and three orders of no species sampling are in red (a). A cladogram depicts the phylogenetic relationships of eight orders, and this tree is taken as one of three rules to exclude the "problematic" gene tree (b).

**Fig. 2.** Phylogenomic trees estimated using ASTRAL and RAxML from 891 orthologous clusters. Strong conflicts are shown for the placement of Zygophyllales. However, RAxML is strongly misled by large amounts of missing data under high gene tree heterogeneity. Bootstrap values are shown for each node as coalescence model/maximum likelihood, and "*" indicates that the clade is supported by bootstrap value 100%. Branch lengths are estimated using maximum likelihood, and scale bar denotes number of substitutions per site.

**Fig. 3.** RF distances among gene trees from 891, 604, 303, 273 and 213 clusters.

**Fig. 4.** RF distances among gene trees from 193 clusters retained after manual curation of the gene trees

**Fig. 5.** Phylogenomic trees inferred using ASTRAL and RAxML from 193 orthologous clusters. Trees with identical topologies are yielded. Bootstrap values are shown for each node as coalescence model/maximum likelihood, and "*" indicates that the clade is supported by bootstrap

value 100%.

**Table 1**. Summary statistics of missing data from five supermatrices for *L. tridentata*.

| | All datasets | # | <10% Missing in each gene | <5% Missing in each gene | <1% Missing in each gene |
|---|---|---|---|---|---|
| Clusters | 891 | 604 | 303 | 273 | 213 |
| Base pair | 1,120,686 | 739,293 | 311,442 | 276,811 | 218,465 |
| % missing sites | 52% | 28% | 2% | 0.7% | 0.2% |
| Position/RAxML | ZM | ZM | ZM | ZM | ZB |
| BS/RAxML | 90% | 95% | 59% | 20% | 45% |
| Position/ASTRAL | ZB | ZB | ZB | ZB | ZB |
| BS/ASTRAL | 100% | 100% | 98% | 100% | 98% |

# denotes that each cluster contains orthologous gene of *L. tridentata*.

ZM denotes that Zygophyllales is sister to Myrtales.

ZB denotes that Zygophyllales is the basal of the “expanded” malvids.

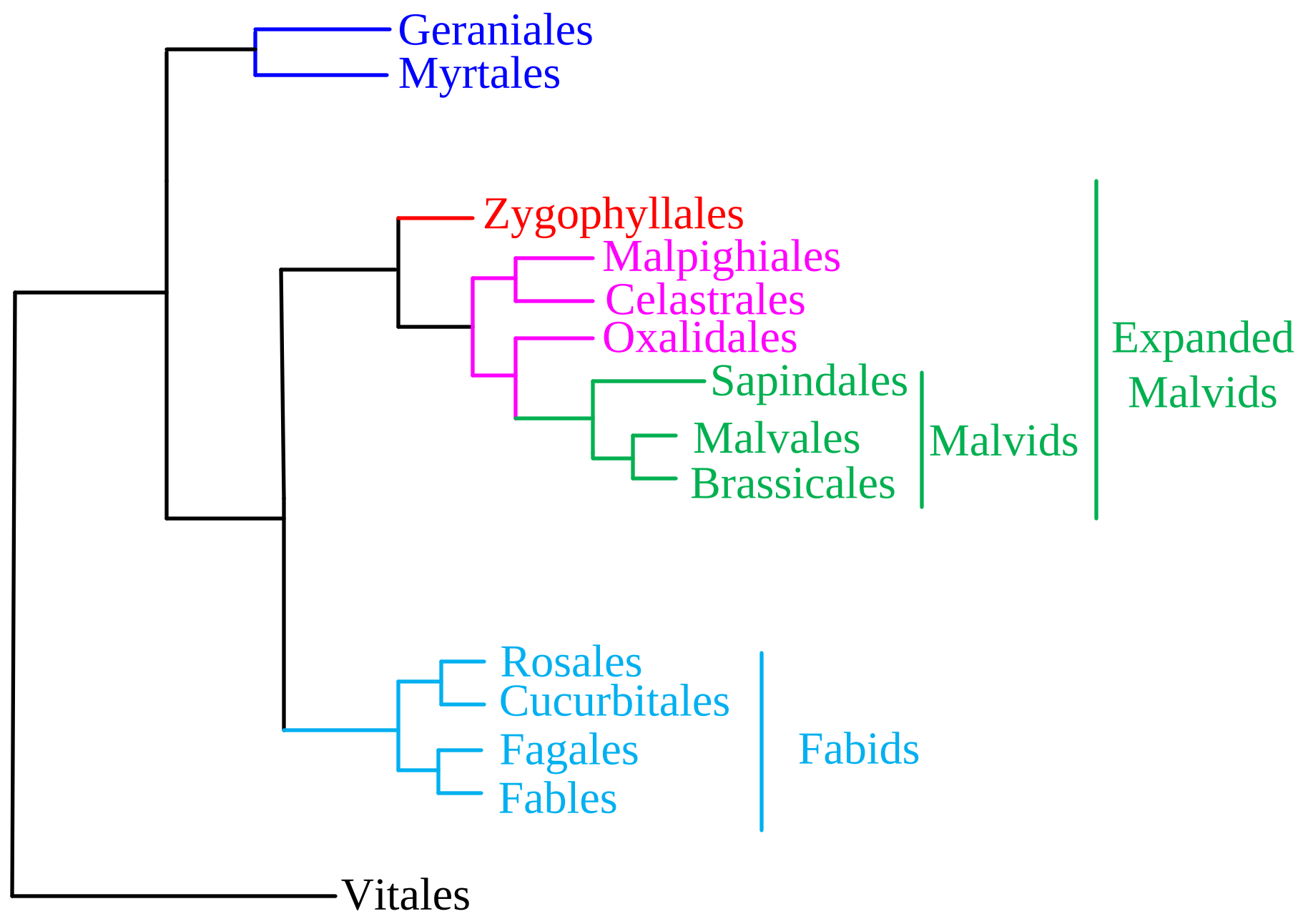

**Highlights:**

The Myrtales-Geraniales clade was basal to the remaining rosids.

The COM clade was not monophyletic, Celastrales and Malpighiales formed to be sister to each other.

The Celastrales-Malpighiales clade and Oxalidales were successive sisters to the malvids.

Zygophyllales was basal to the “expanded” malvids (including COM orders).

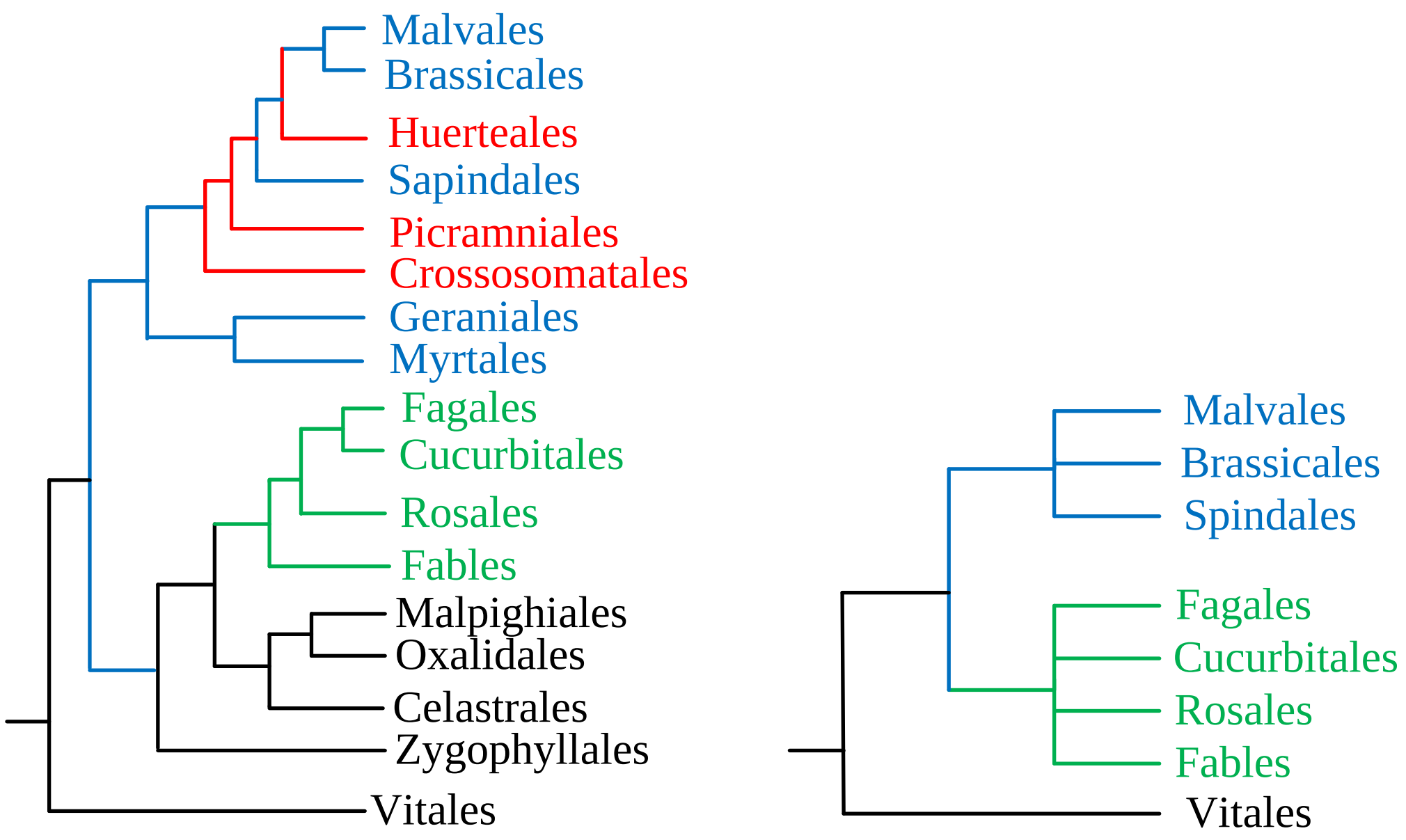

(a) (b)

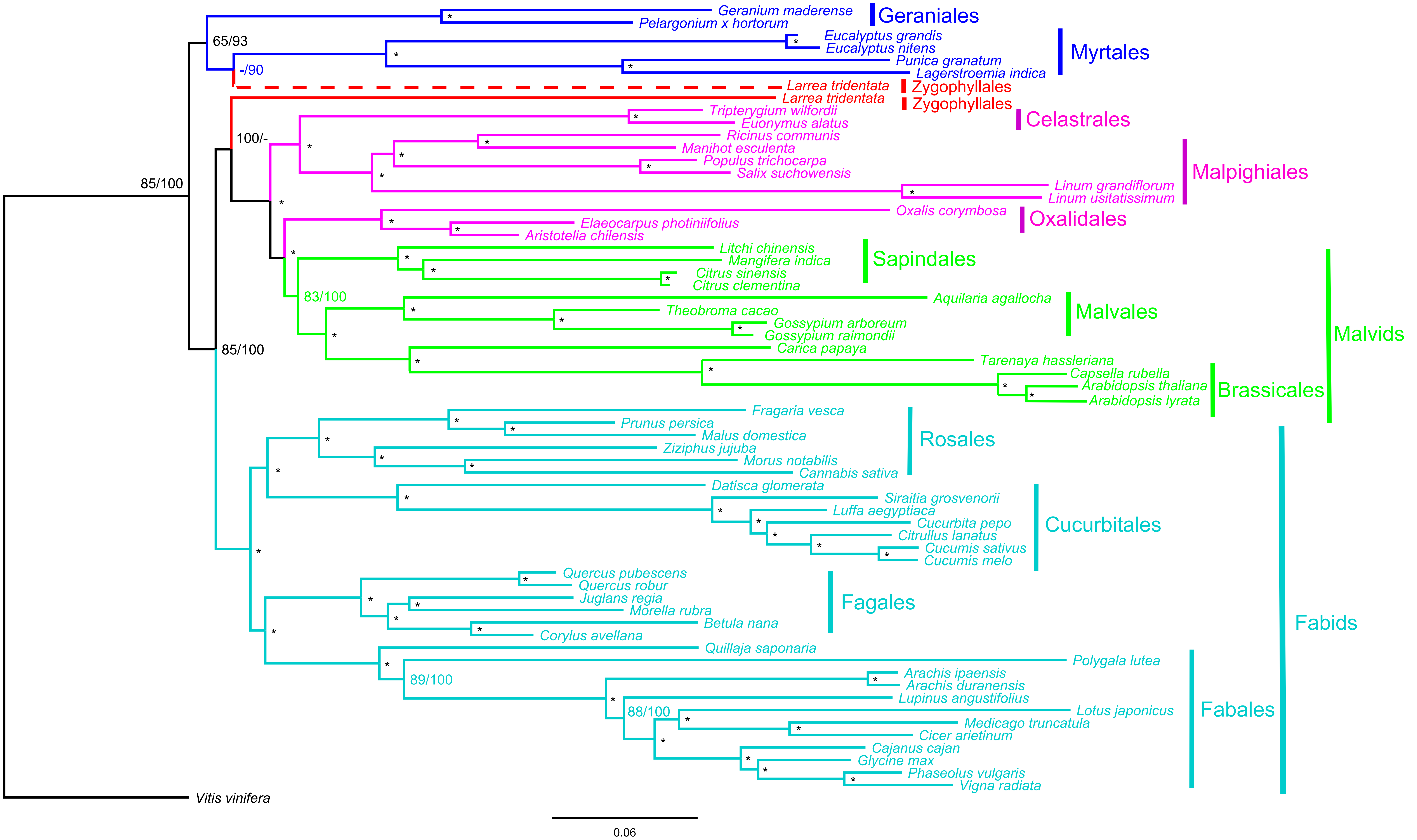

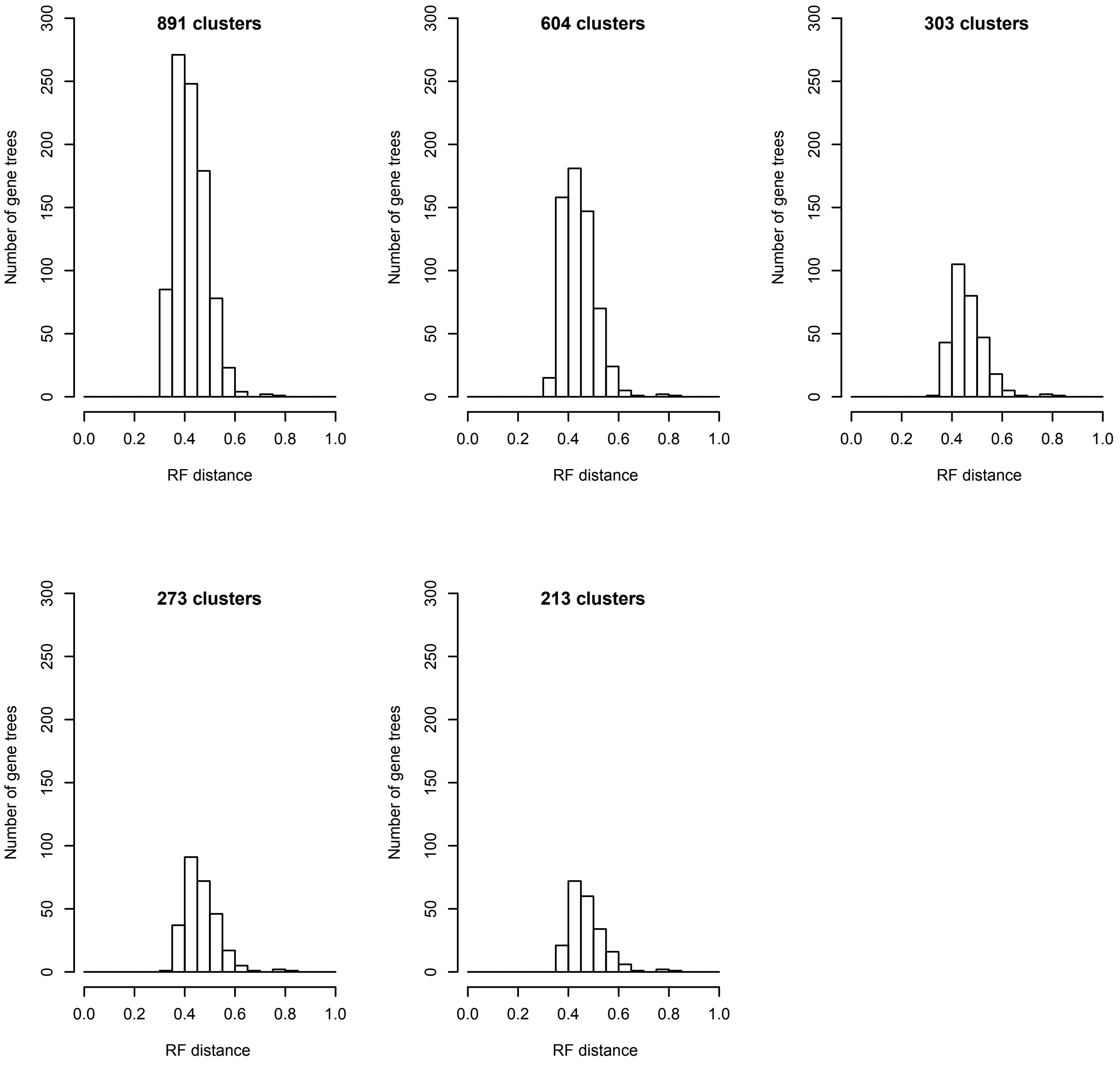
891 clusters
604 clusters
303 clusters
273 clusters
213 clusters
Number of gene trees
RF distance
0 50 100 150 200 250 300
0.0 0.2 0.4 0.6 0.8 1.0

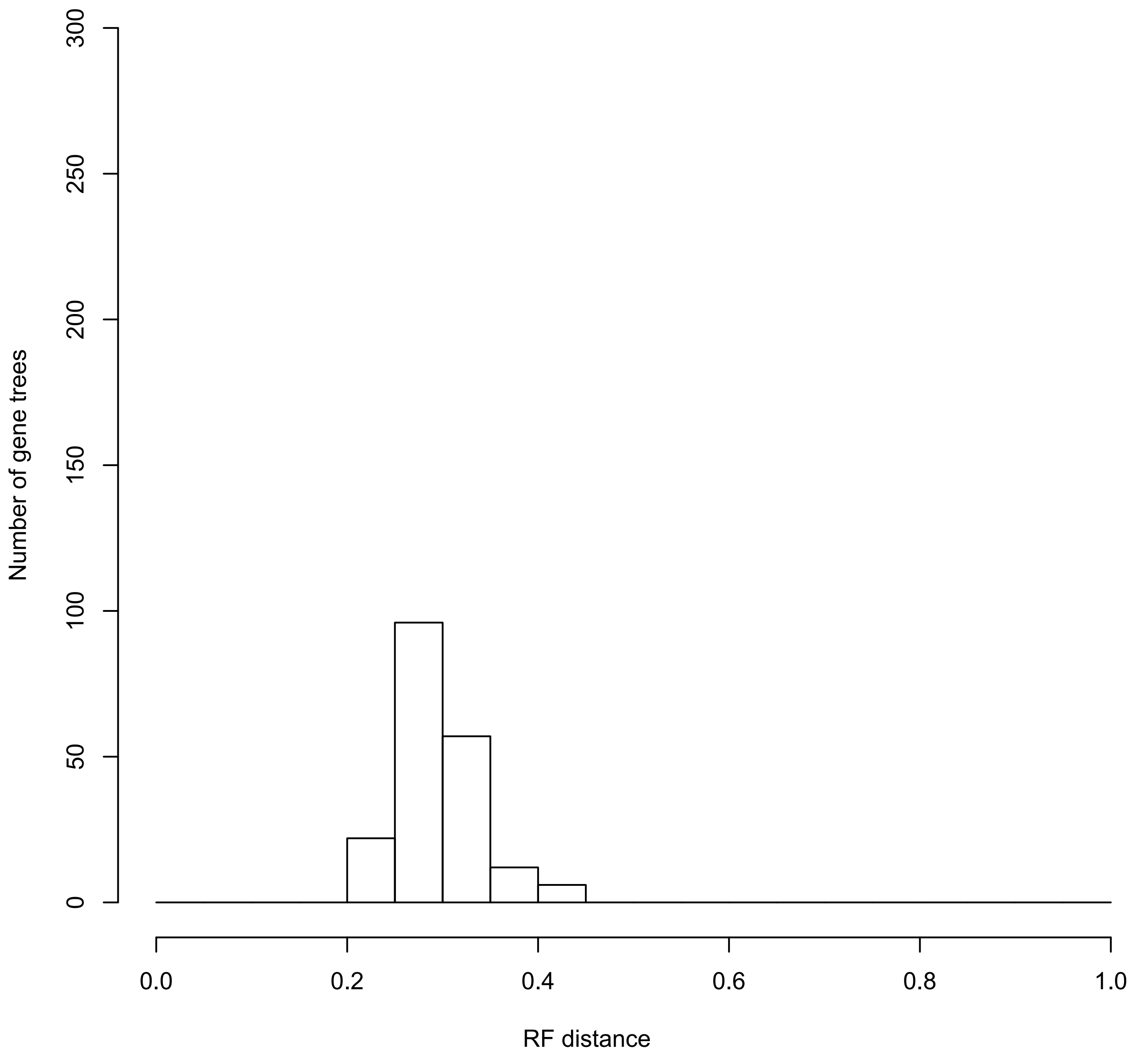

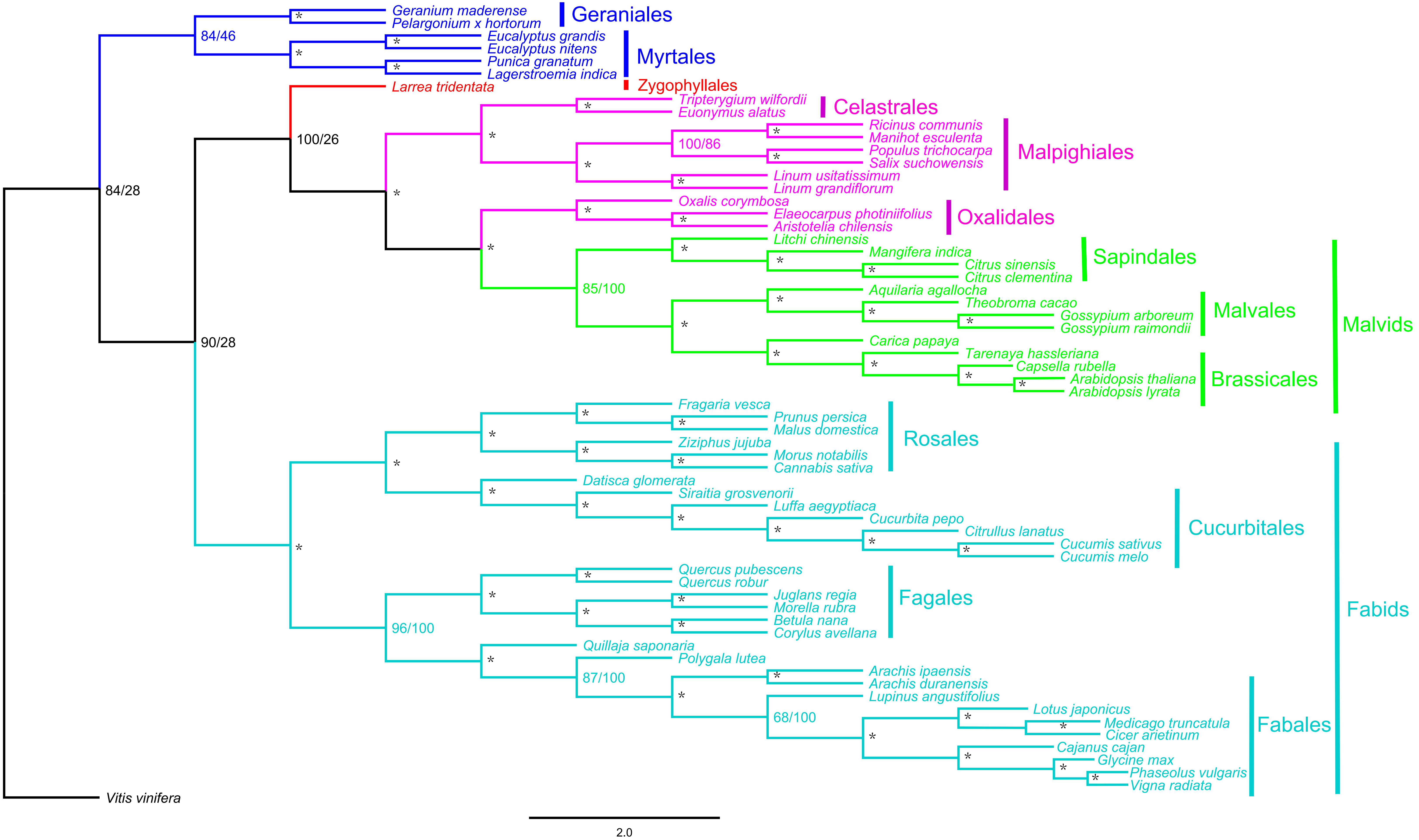

# Supplementary material

**Table S1**. Overview of genes and species used in this study.

**Figure S1.** Data flow diagram of bioinformatics pipeline.

**Figure S2**. Maximum-likelihood tree estimated in RAxML from 891 clusters, 63 species based on only the first and second codon positions (Matrix B).

**Figure S3**. Coalescent-based tree estimated in ASTRAL from 891 orthologous clusters based on the first and second codon positions.

**Figure S4**. Maximum-likelihood trees estimated in RAxML from 891 clusters, 63 species based on the fast sites (Matrix C).

**Figure S5**. Coalescent-based tree is estimated in ASTRAL from 891 orthologous clusters based on the fast sites.

**Figure S6**. Maximum-likelihood tree is estimated in RAxML from 891 clusters, 63 species based on the slow sites. (Matrix D).

**Figure S7**. Coalescent-based tree is estimated in ASTRAL from 891 orthologous clusters based on the slow sites.

**Figure S8**. Maximum-likelihood tree is estimated in RAxML from 891 clusters, 57 species based all codon positions (Matrix E).

**Figure S9**. Maximum-likelihood tree is estimated in RAxML from 604 clusters, 63 species.

**Figure S10**. Maximum-likelihood tree is estimated in RAxML from 303 clusters, 63 species.

**Figure S11**. Maximum-likelihood tree is estimated in RAxML from 273 clusters, 63 species.

**Figure S12**. Maximum-likelihood tree is estimated in RAxML from 213 clusters, 63 species.

**Figure S13**. Coalescent species tree is estimated in ASTRAL4.7.7 from 604 clusters, 63 species.

**Figure S14**. Coalescent species tree is estimated in ASTRAL4.7.7 from 303 clusters, 63 species.

**Figure S15**. Coalescent species tree is estimated in ASTRAL4.7.7 from 273 clusters, 63 species.

**Figure S16**. Coalescent species tree is estimated in ASTRAL4.7.7 from 213 clusters, 63 species.

**Figure S17**. Single gene tree (RAxML_bipartitions.431154_translatorx_nt_trimal) is estimated in RAxML from the “431154” cluster.

**Figure S18**. Single gene tree (RAxML_bipartitions.431249_translatorx_nt_trimal) is estimated in RAxML from the “431249” cluster.

**Figure S19**. Single gene tree (RAxML_bipartitions.431263_translatorx_nt_trimal) is estimated in RAxML from the “431263” cluster.

| Species | Abbreviation | Family | Order |
|---|---|---|---|
| *Cucumis sativus* | Csat | Cucurbitaceae | Cucurbitales |
| *Cucumis melo* | Cmelo | Cucurbitaceae | Cucurbitales |
| *Citrullus lanatus* | Clan | Cucurbitaceae | Cucurbitales |
| *Siraitia grosvenorii* | Sirgro | Cucurbitaceae | Cucurbitales |
| *Cucurbita pepo* | Cpepo | Cucurbitaceae | Cucurbitales |
| *Luffa aegyptiaca* | Lufaeg | Cucurbitaceae | Cucurbitales |
| *Datisca glomerata* | Dglo | Datiscacea | Cucurbitales |
| *Betula nana* | Bnana | Betulaceae | Fagales |
| *Corylus avellana* | Cave | Betulaceae | Fagales |
| *Quercus robur* | Qrob | Fagaceae | Fagales |
| *Quercus pubescens* | Qpubes | Fagaceae | Fagales |
| *Morella rubra* | Mrubra | Myricaceae | Fagales |
| *Juglans regia* | Jregia | Juglandaceae | Fagales |
| *Prunus persica* | Ppersica | Rosaceae | Rosales |
| *Fragaria vesca* | Fves | Rosaceae | Rosales |
| *Malus domestica* | Mdomes | Rosaceae | Rosales |
| *Morus notabilis* | Mnot | Moraceae | Rosales |
| *Cannabis sativa* | Cansat | Cannabaceae | Rosales |
| *Ziziphus jujuba* | Zizjuj | Rhamnaceae | Rosales |
| *Polygala lutea* | Plutea | Polygalaceae | Fabales |
| *Quillaja saponaria* | Qsapon | Quillajaceae | Fabales |
| *Glycine max* | glyma | Fabaceae | Fabales |
| *Medicago truncatula* | medtr | Fabaceae | Fabales |
| *Phaseolus vulgaris* | Pvul | Fabaceae | Fabales |
| *Cajanus cajan* | Ccajan | Fabaceae | Fabales |
| *Cicer arietinum* | Carietin | Fabaceae | Fabales |
| *Lotus japonicus* | Ljapon | Fabaceae | Fabales |
| *Arachis duranensis* | Aradur | Fabaceae | Fabales |
| *Arachis ipaensis* | Araipa | Fabaceae | Fabales |
| *Lupinus angustifolius* | Lupangu | Fabaceae | Fabales |
| *Vigna radiata* | Virad | Fabaceae | Fabales |
| *Euonymus alatus* | Eala | Celastraceae | Celastrales |
| *Tripterygium wilfordii* | Twil | Celastraceae | Celastrales |
| *Oxalis corymbosa* | Ocor | Oxalidaceae | Oxalidales |
| *Aristotelia chilensis* | Achilen | Elaeocarpaceae | Oxalidales |
| *Elaeocarpus photiniifolius* | Ephot | Elaeocarpaceae | Oxalidales |
| *Larrea tridentata* | Ltri | Zygophyllaceae | Zygophyllales |
| *Populus trichocarpa* | poptr | Salicaceae | Malpighiales |
| *Salix suchowensis* | Ssuc | Salicaceae | Malpighiales |
| *Ricinus communis* | Rcommunis | Euphorbiaceae | Malpighiales |
| *Manihot esculenta* | Mesu | Euphorbiaceae | Malpighiales |
| *Linum usitatissimum* | Lusi | Linaceae | Malpighiales |
| *Linum grandiflorum* | Lgrand | Linaceae | Malpighiales |
| *Geranium maderense* | Gmad | Geraniaceae | Geraniales |
| *Pelargonium x hortorum* | Phor | Geraniaceae | Geraniales |
| *Eucalyptus grandis* | Egra | Myrtaceae | Myrtales |
| *Eucalyptus nitens* | Enit | Myrtaceae | Myrtales |
| *Lagerstroemia indica* | Lindica | Lythraceae | Myrtales |
| *Punica granatum* | Pgran | Lythraceae | Myrtales |
| *Citrus clementina* | Ccle | Rutaceae | Sapindales |
| *Citrus sinensis* | Csin | Rutaceae | Sapindales |
| *Mangifera indica* | Mind | Anacardiaceae | Sapindales |
| *Litchi chinensis* | Litchi | Sapindaceae | Sapindales |
| *Aquilaria agallocha* | Aagallo | Thymelaeaceae | Malvales |
| *Theobroma cacao* | Tcacao | Sterculiaceae | Malvales |
| *Gossypium raimondii* | Grai | Malvaceae | Malvales |
| *Gossypium arboreum* | Garboreum | Malvaceae | Malvales |
| *Tarenaya hassleriana* | Thas | Cleomaceae | Brassicales |
| *Carica papaya* | Cpap | Caricaceae | Brassicales |
| *Arabidopsis thaliana* | arath | Brassicaceae | Brassicales |
| *Arabidopsis lyrata* | Alyr | Brassicaceae | Brassicales |
| *Capsella rubella* | Crub | Brassicaceae | Brassicales |
| *Vitis vinifera* | vitvi | Vitaceae | Vitales |

| Type | Database | Gene Numbers |
| --- | --- | --- |
| genome | Phytozome | 30364 |
| genome | Phytozome | 34848 |
| genome | Phytozome | 23440 |
| Transcriptome(Illumina) | NCBI | 44106 |
| Transcriptome(Illumina) | NCBI | 28798 |
| Transcriptome(Illumina) | NCBI | 65262 |
| Transcriptome(Illumina)+454 | NCBI | 77233 |
| genome | NCBI | 59194 |
| Transcriptome(Illumina)+EST | NCBI | 19992 |
| Transcriptome(Illumina) | NCBI | 30114 |
| Transcriptome(Illumina) | NCBI | 19141 |
| Transcriptome(Illumina) | NCBI | 26572 |
| Transcriptome(Illumina)+454 | NCBI | 21849 |
| genome | Phytozome | 28701 |
| genome | Phytozome | 32831 |
| genome | Phytozome | 63517 |
| genome | NCBI | 21888 |
| genome | NCBI | 30074 |
| genome | NCBI | 29051 |
| Transcriptome(Illumina)+454 | NCBI | 29333 |
| Transcriptome(Illumina)+454 | NCBI | 22912 |
| genome | Phytozome | 73320 |
| genome | Phytozome | 45888 |
| genome | Phytozome | 31638 |
| genome | gigadb.org | 48680 |
| genome | gigadb.org | 28269 |
| genome | Phytozome | 38482 |
| genome | NCBI | 37842 |
| genome | NCBI | 38967 |
| Transcriptome(Illumina) | NCBI | 32592 |
| genome | NCBI | 22368 |
| Transcriptome(454)+EST | NCBI | 15527 |
| Transcriptome(454)+EST | NCBI | 81820 |
| Transcriptome(Illumina) | This study | 47627 |
| Transcriptome(Illumina) | NCBI | 20694 |
| Transcriptome(454) | NCBI | 8494 |
| Transcriptome(Illumina)+EST | NCBI | 18703 |
| genome | Phytozome | 73013 |
| genome | Phytozome | 26599 |
| genome | Phytozome | 31221 |
| genome | Phytozome | 43151 |
| genome | Phytozome | 43484 |
| Transcriptome(Illumina) | NCBI | 35135 |
| Transcriptome(Illumina,454) | NCBI | 51922 |
| Transcriptome(Illumina,454) | NCBI | 48671 |
| genome | Phytozome | 46315 |
| Transcriptome(Illumina) | NCBI | 104099 |
| Transcriptome(Illumina) | NCBI | 29497 |
| Transcriptome(Illumina) | NCBI | 25201 |
| genome | Phytozome | 33929 |
| genome | Phytozome | 46147 |
| Transcriptome(Illumina) | NCBI | 23943 |
| Transcriptome(Illumina) | NCBI | 20135 |
| Transcriptome(Illumina) | NCBI | 21687 |
| genome | Phytozome | 44404 |
| genome | Phytozome | 77267 |
| genome | Phytozome | 40134 |
| genome | Phytozome | 28917 |
| genome | Phytozome | 27760 |
| genome | Phytozome | 35380 |
| genome | Phytozome | 32670 |
| genome | Phytozome | 28447 |
| genome | Phytozome | 26346 |

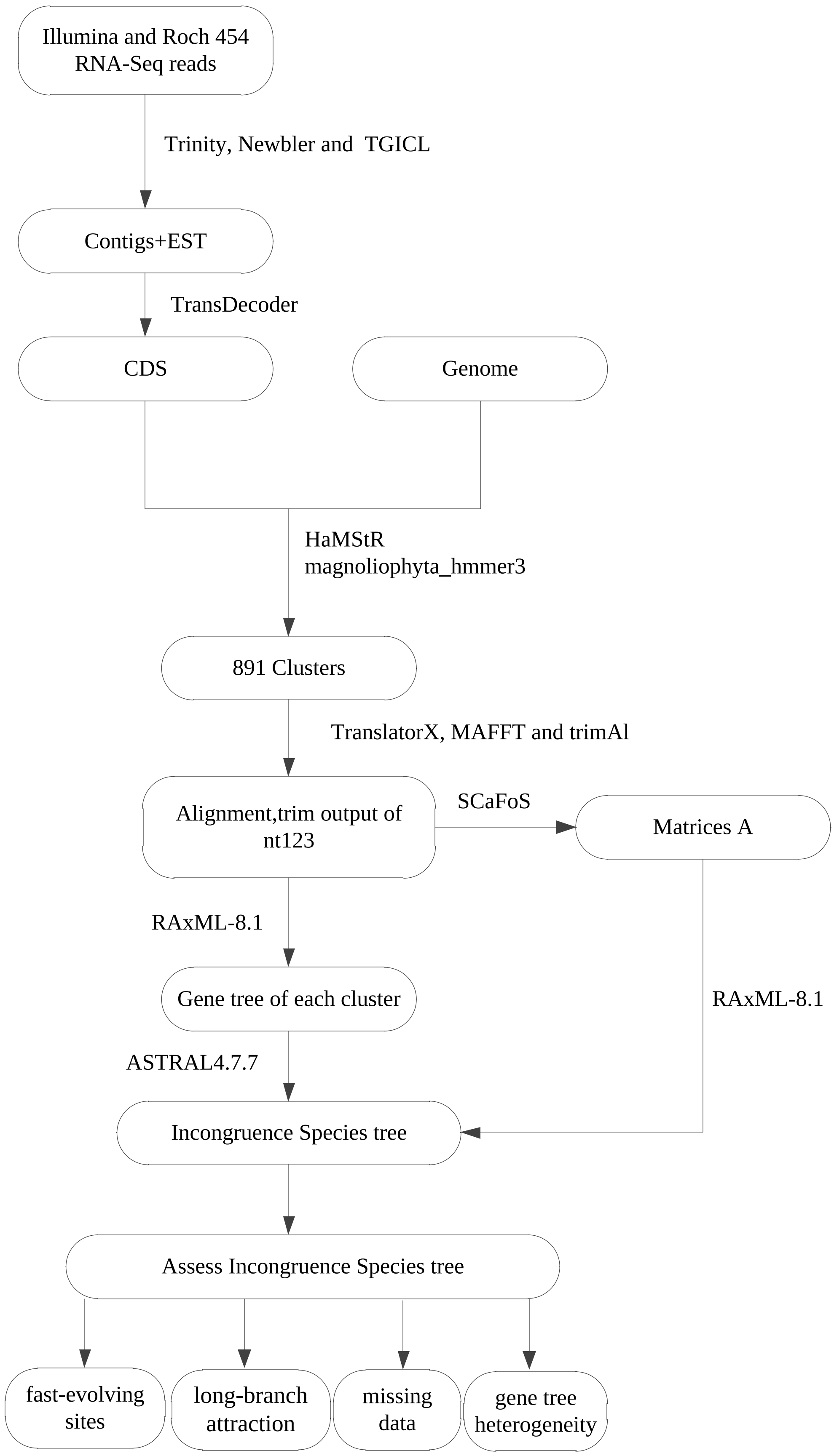
Illumina and Roch 454 RNA-Seq reads
Trinity, Newbler and TGICL
Contigs+EST
TransDecoder
CDS
Genome
HaMStR
magnoliophyta_hmmer3
891 Clusters
TranslatorX, MAFFT and trimAl
Alignment,trim output of nt123
SCaFoS
Matrices A
RAxML-8.1
Gene tree of each cluster
RAxML-8.1
ASTRAL4.7.7
Incongruence Species tree
Assess Incongruence Species tree
fast-evolving sites
long-branch attraction
missing data
gene tree heterogeneity

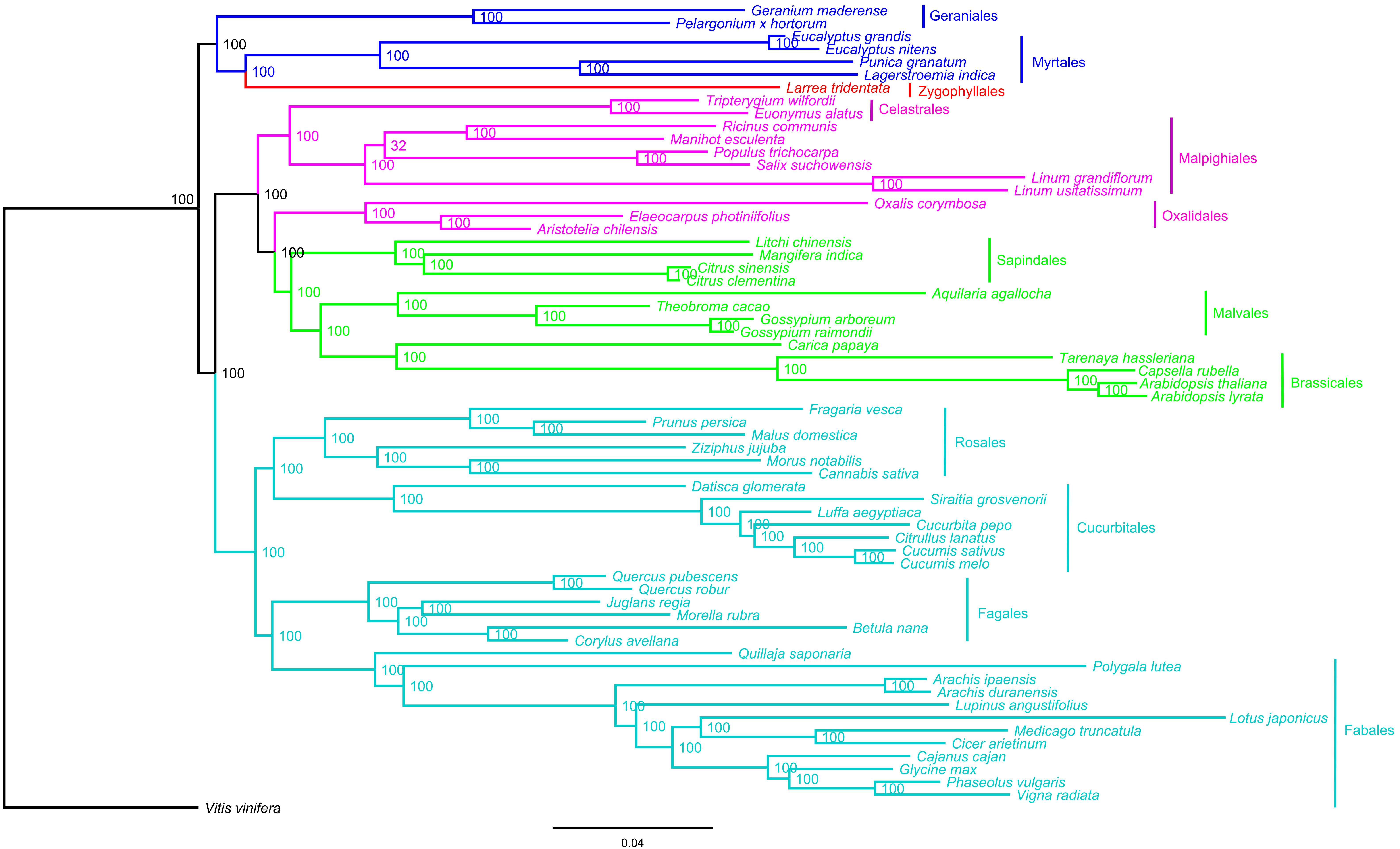

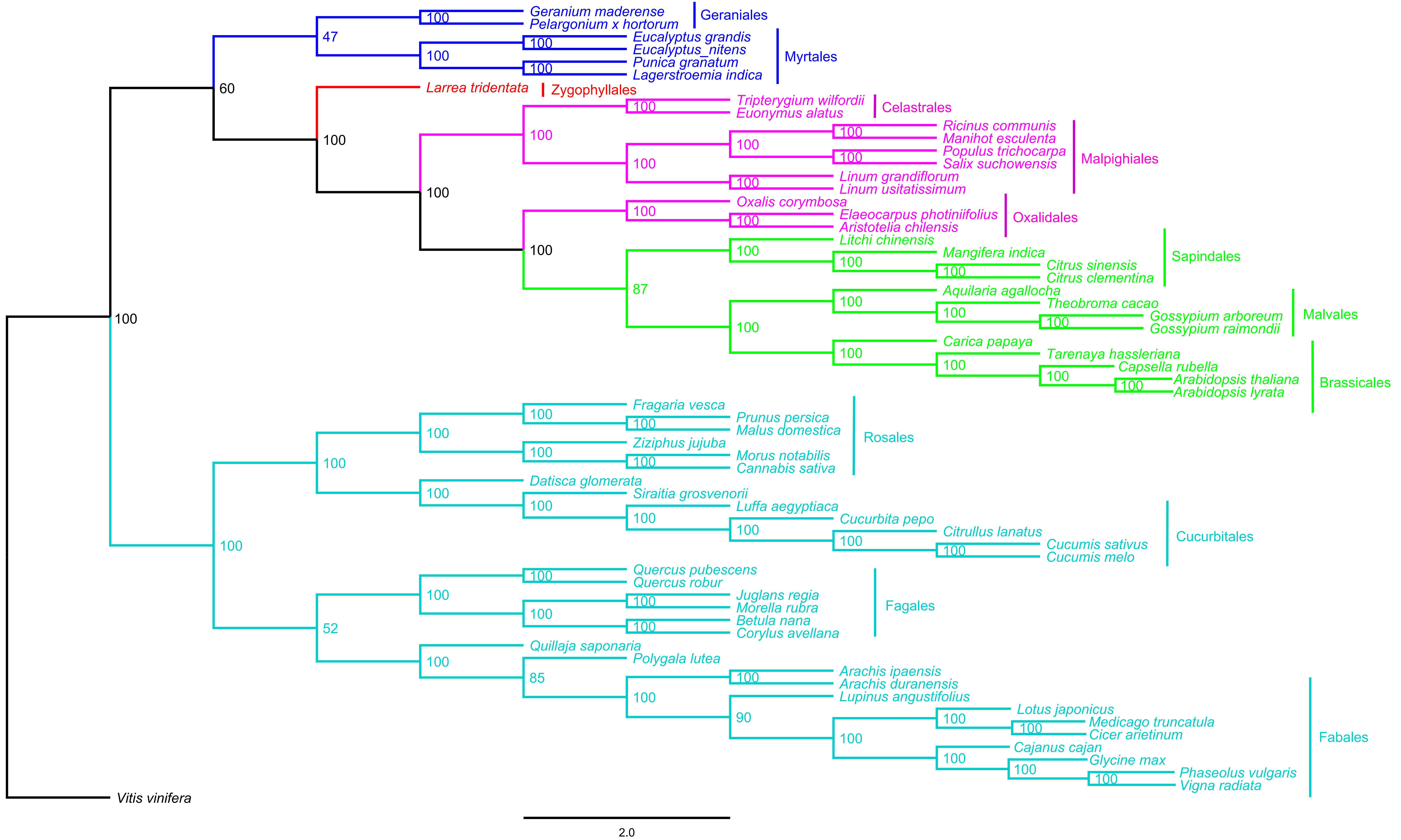

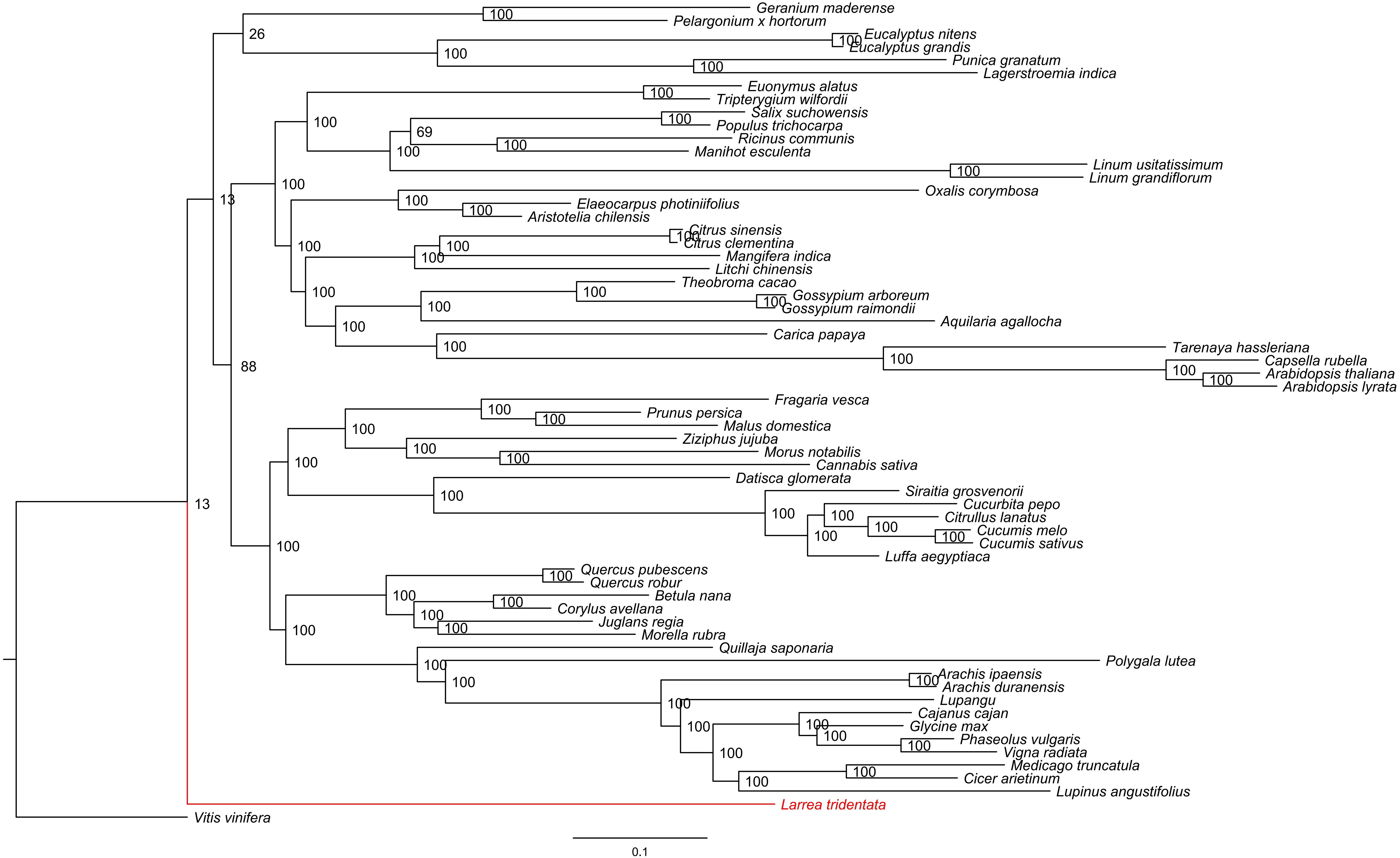

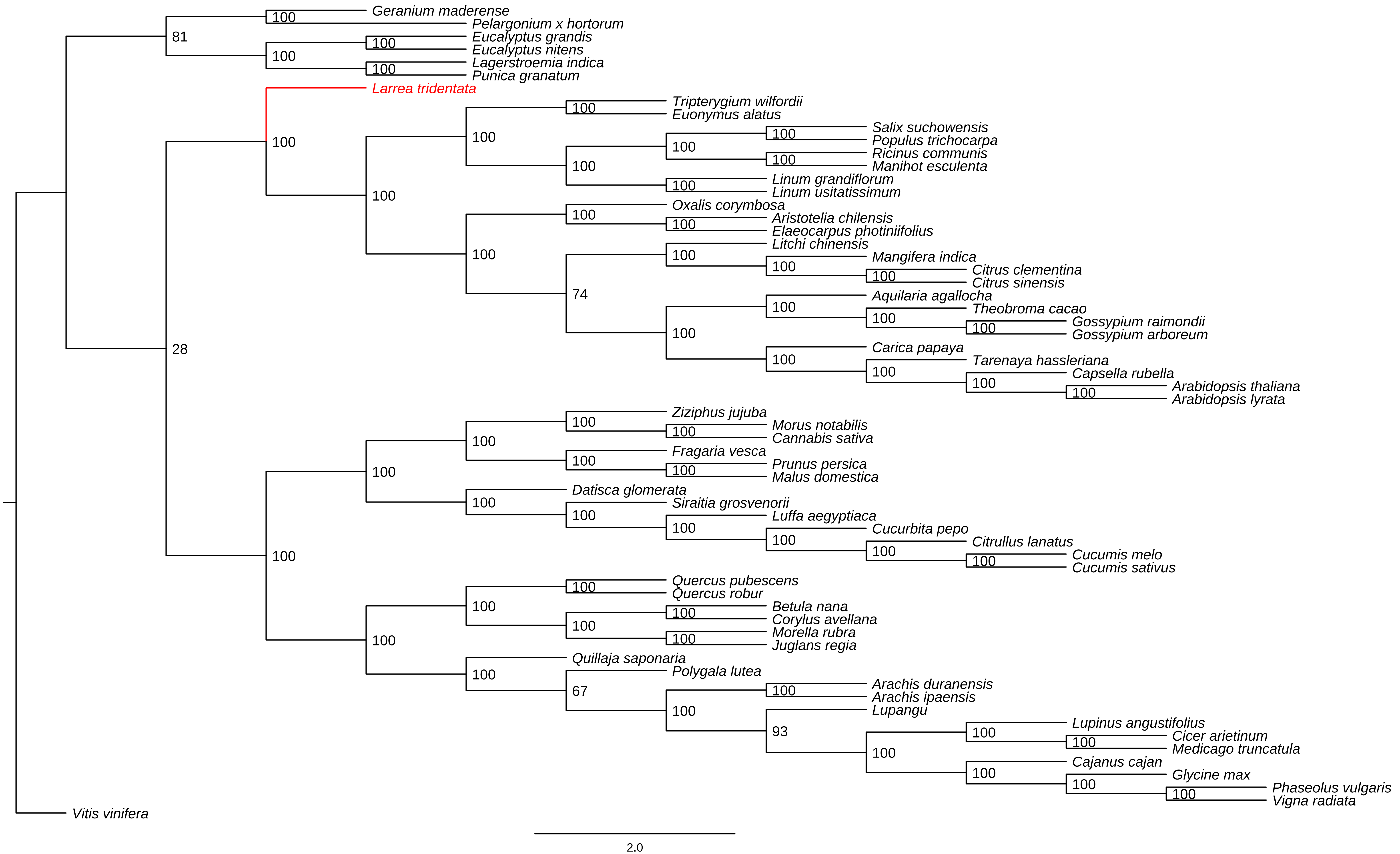

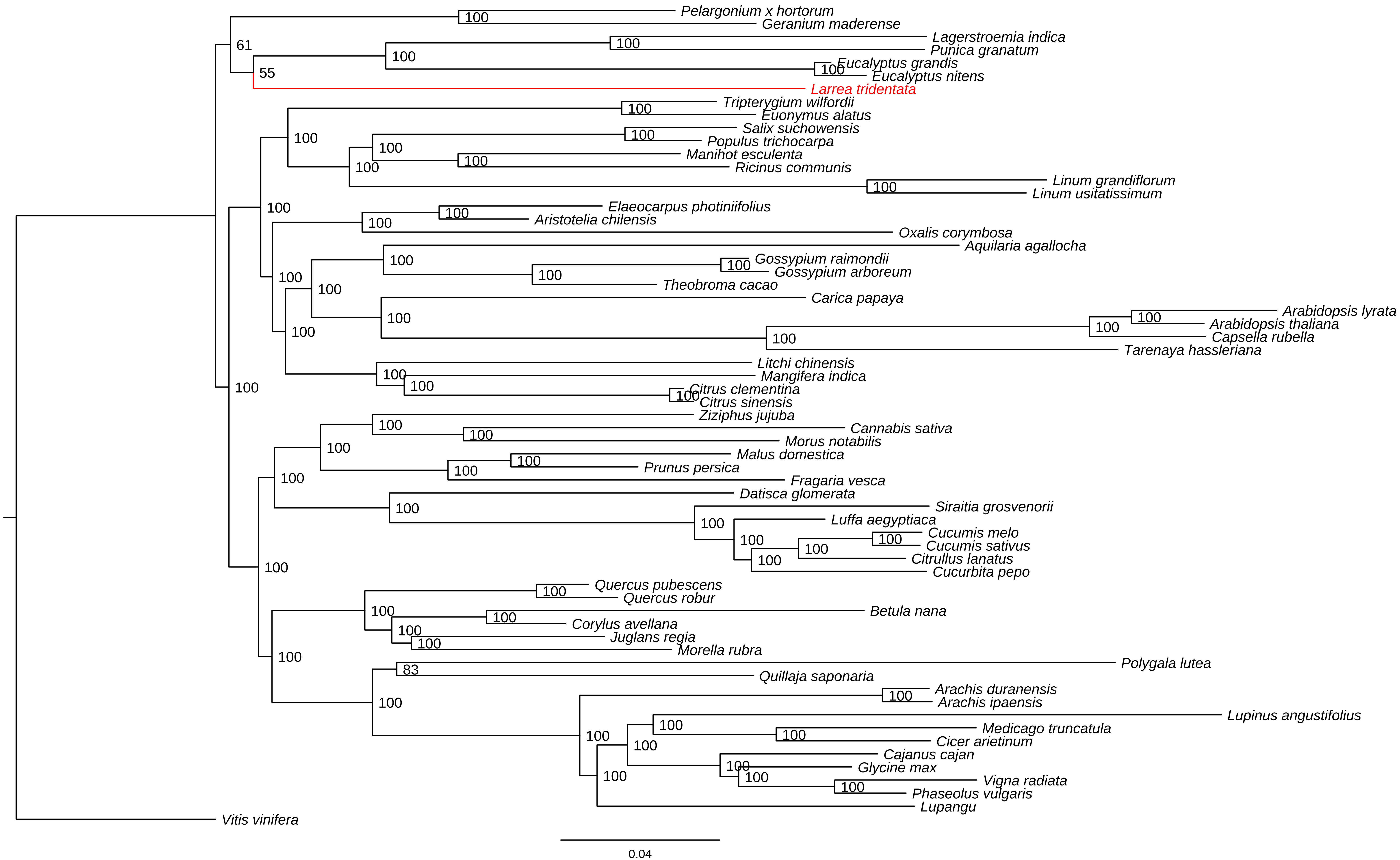

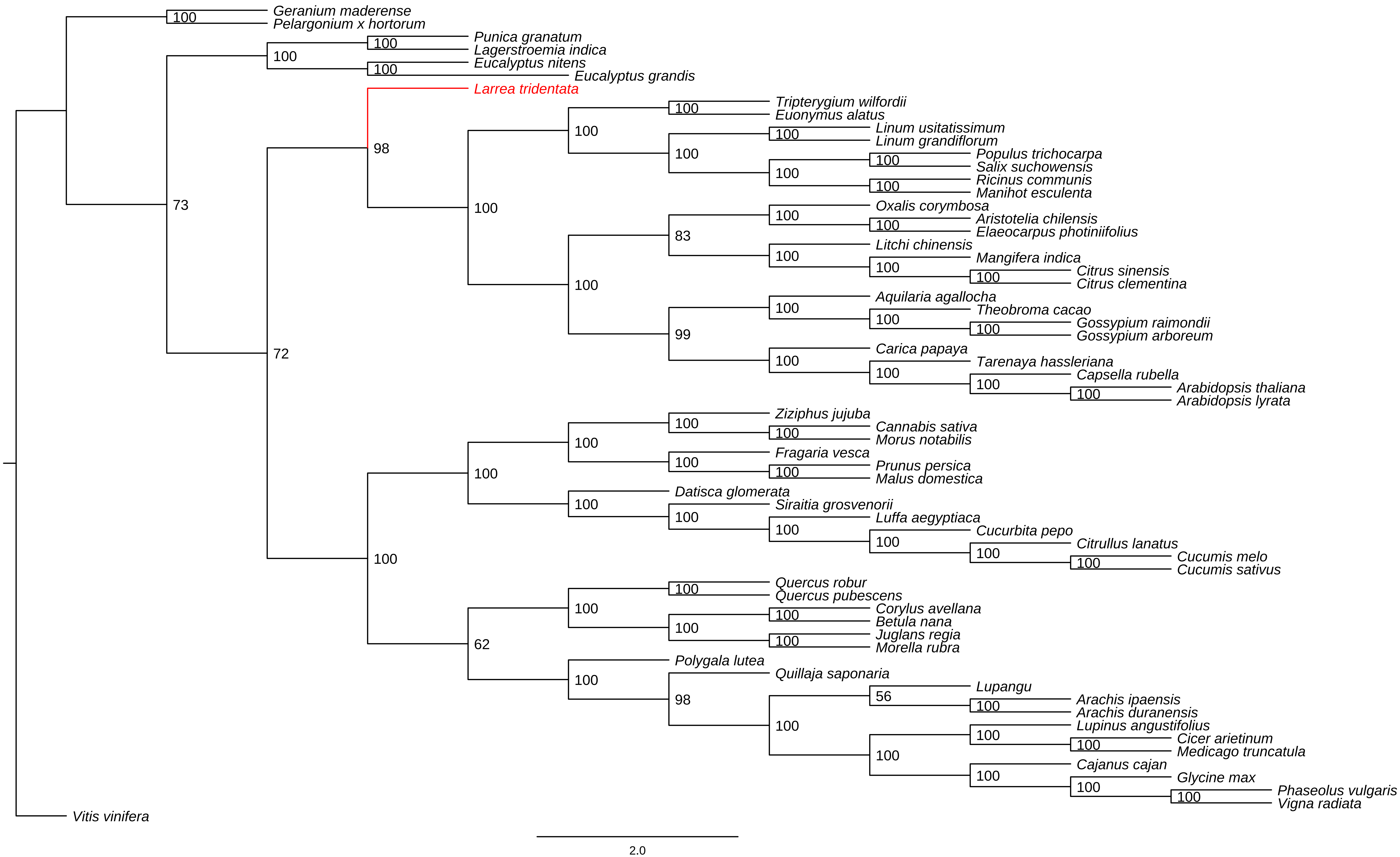

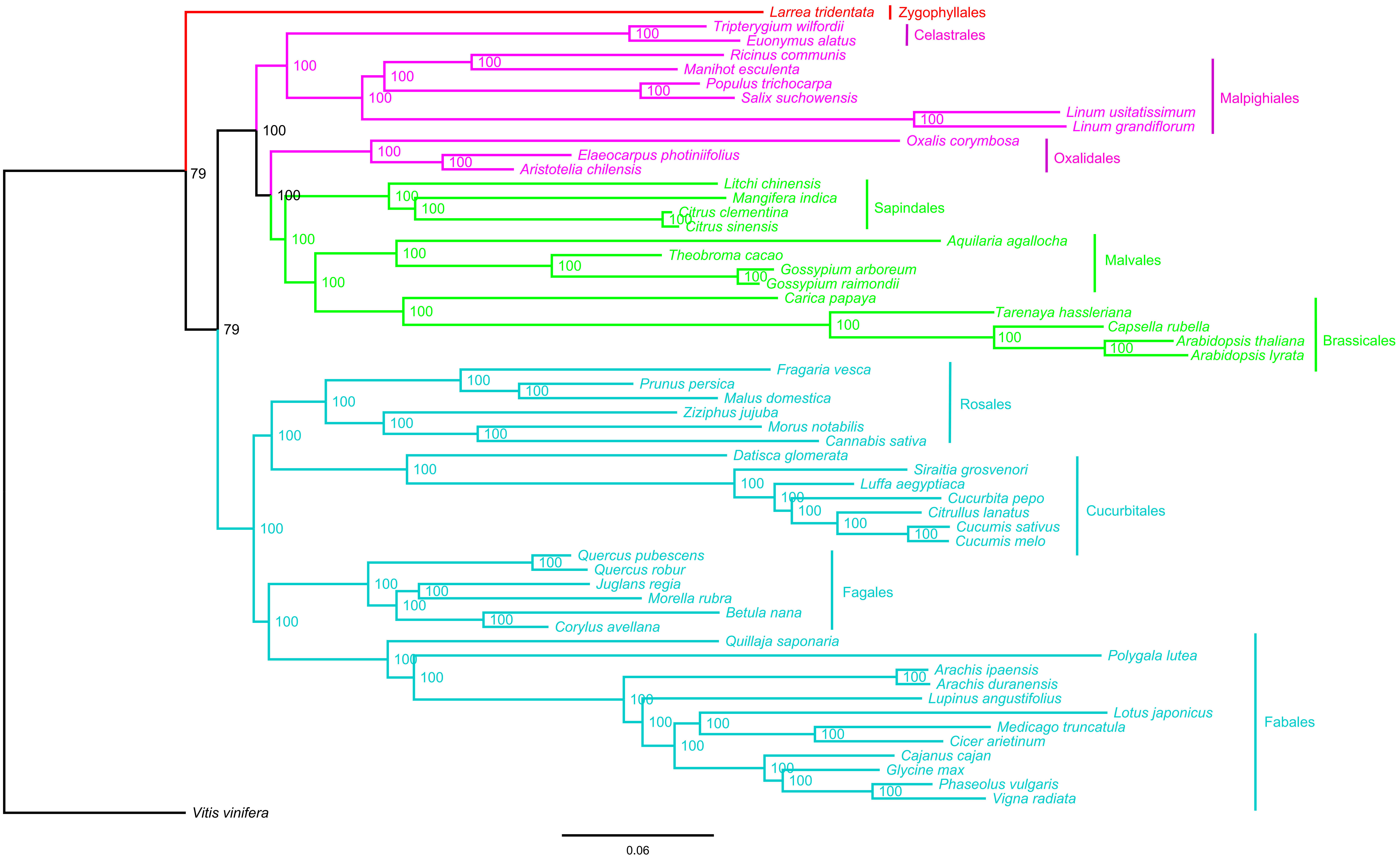

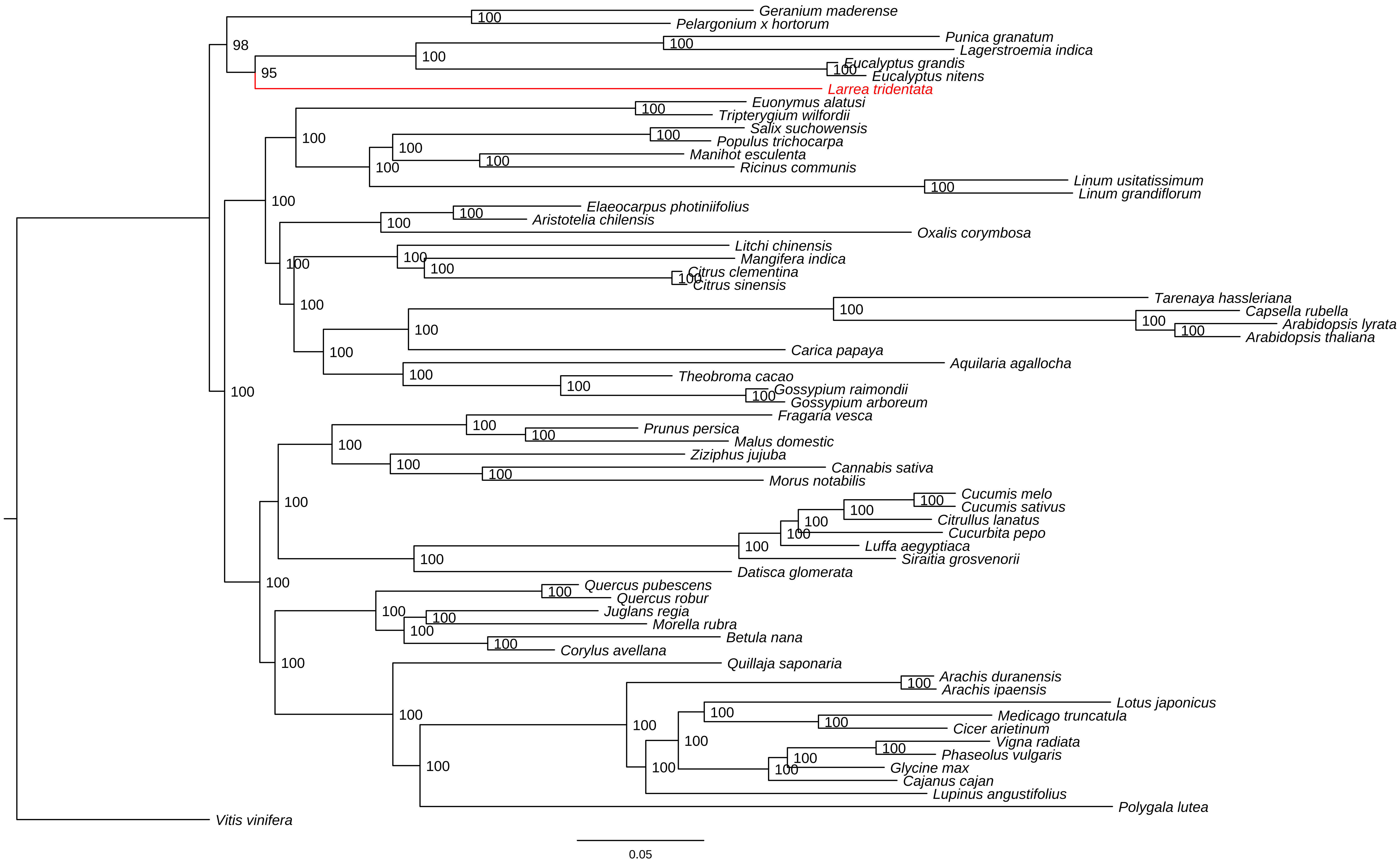

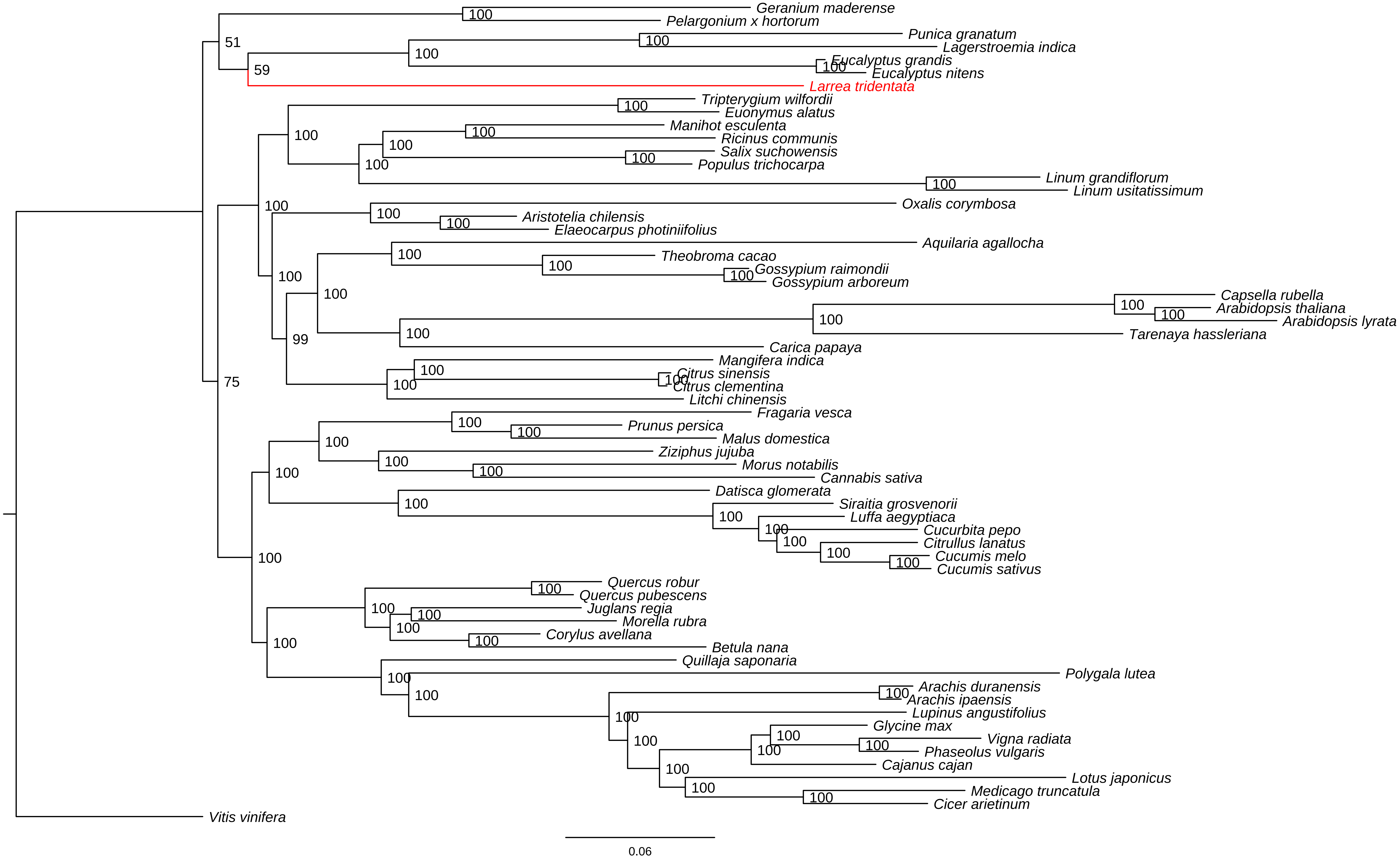

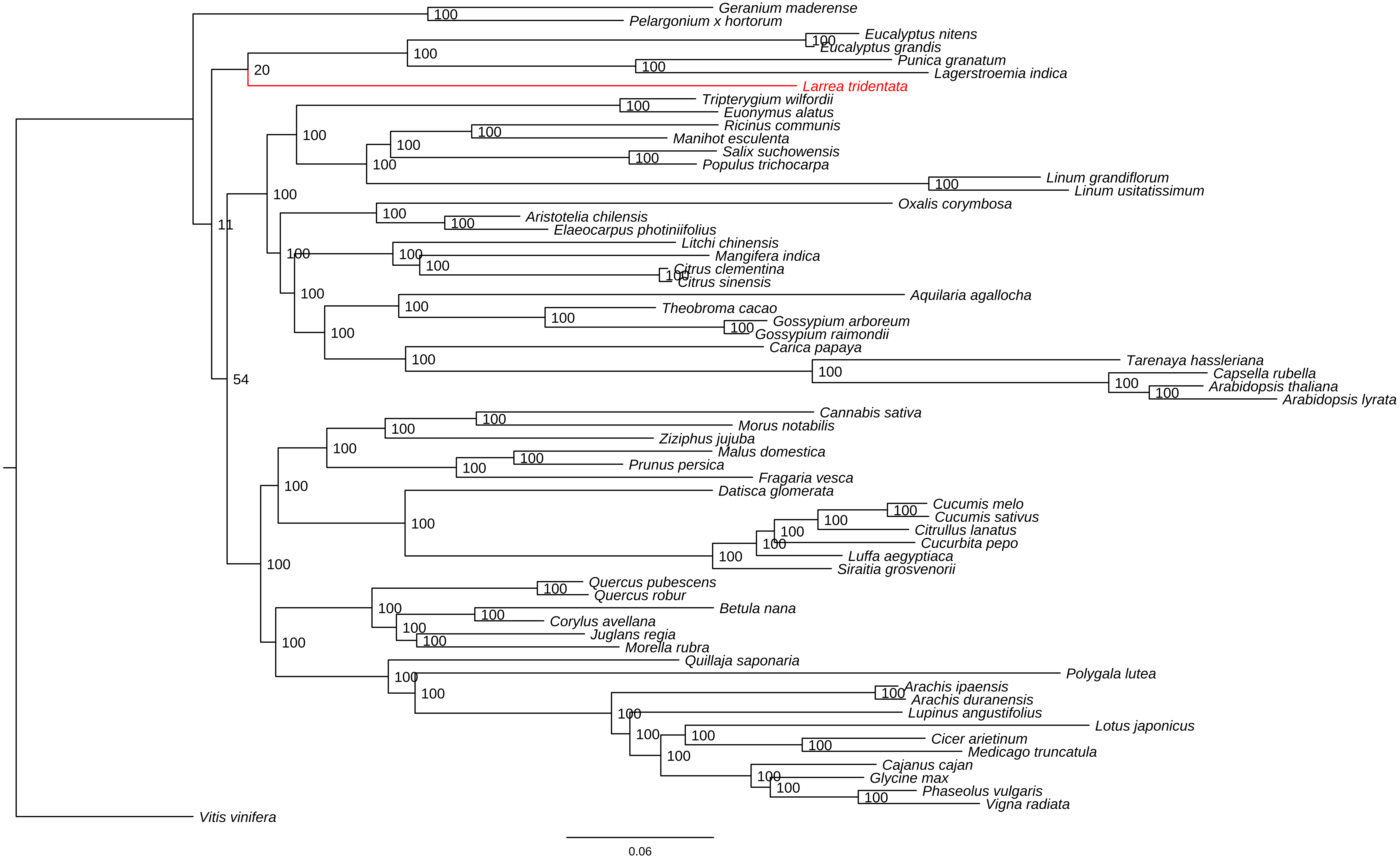

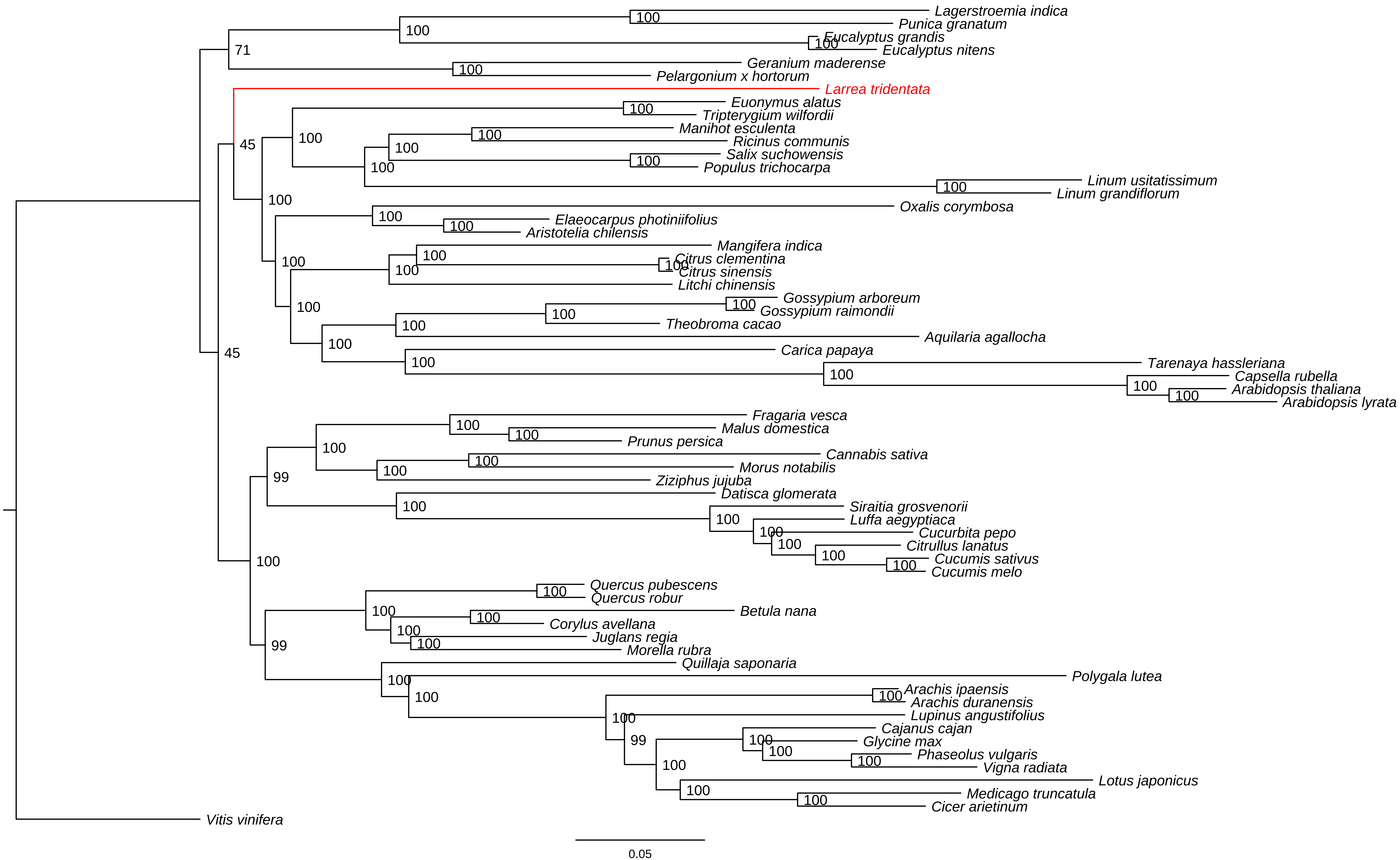

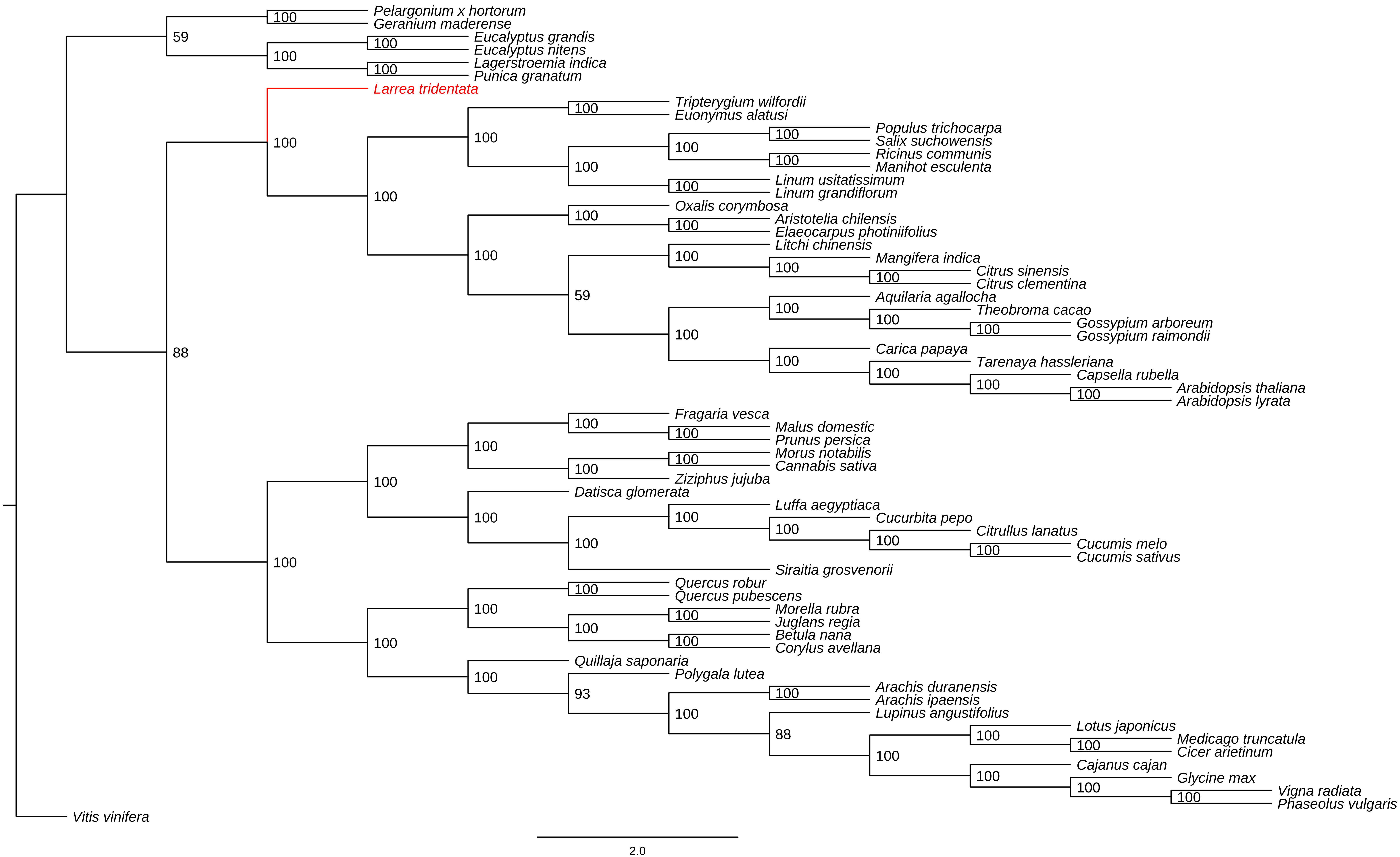

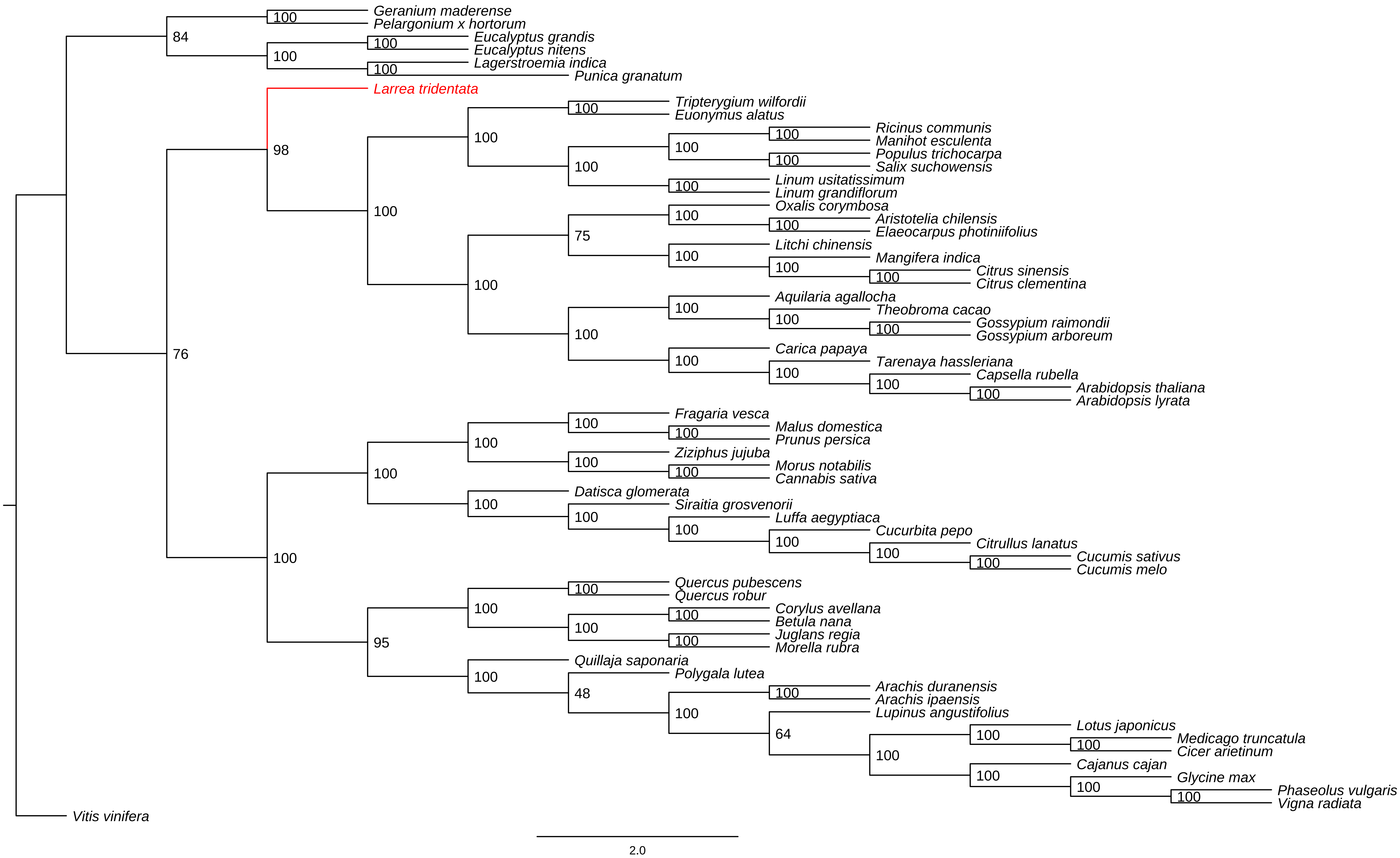

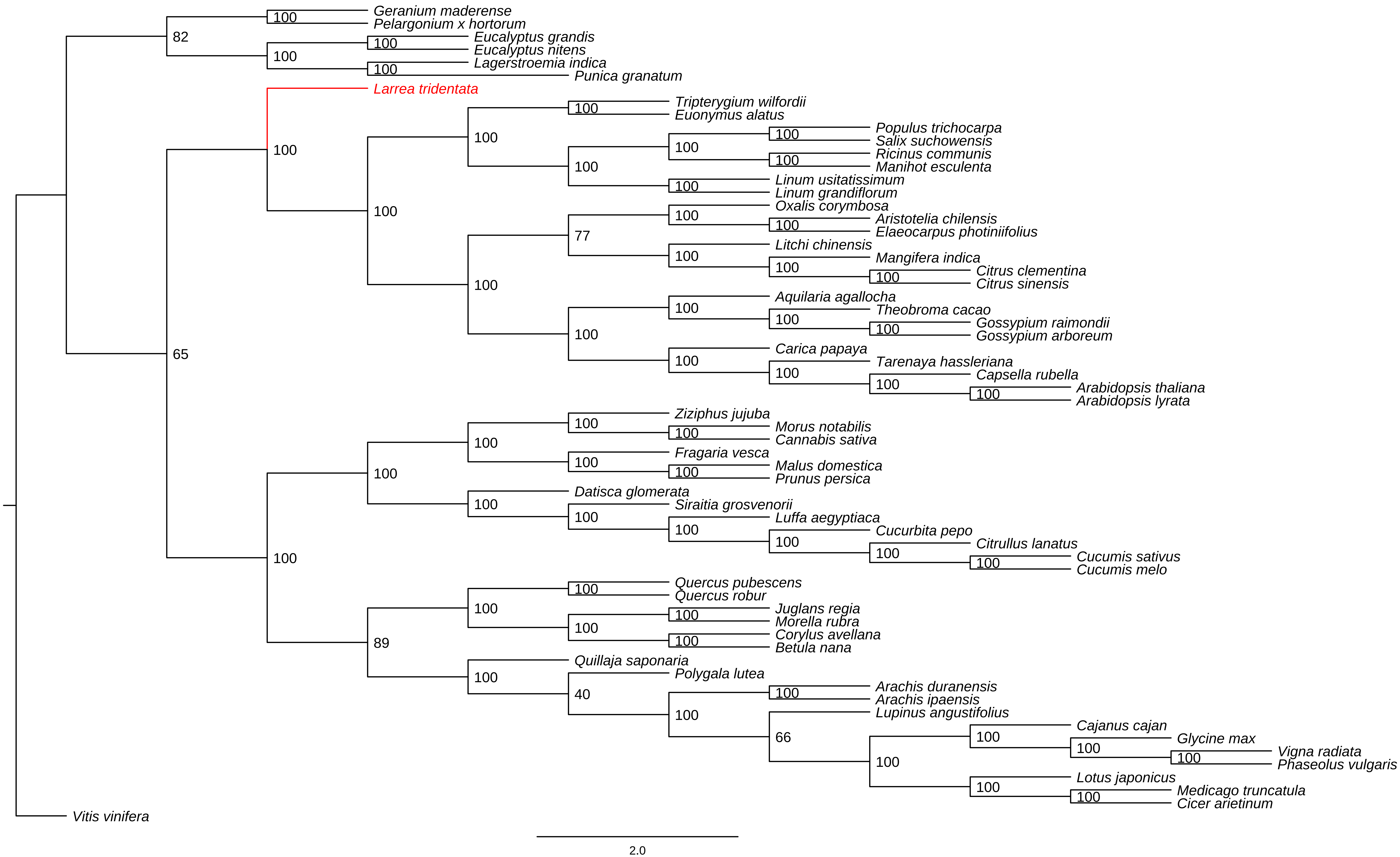

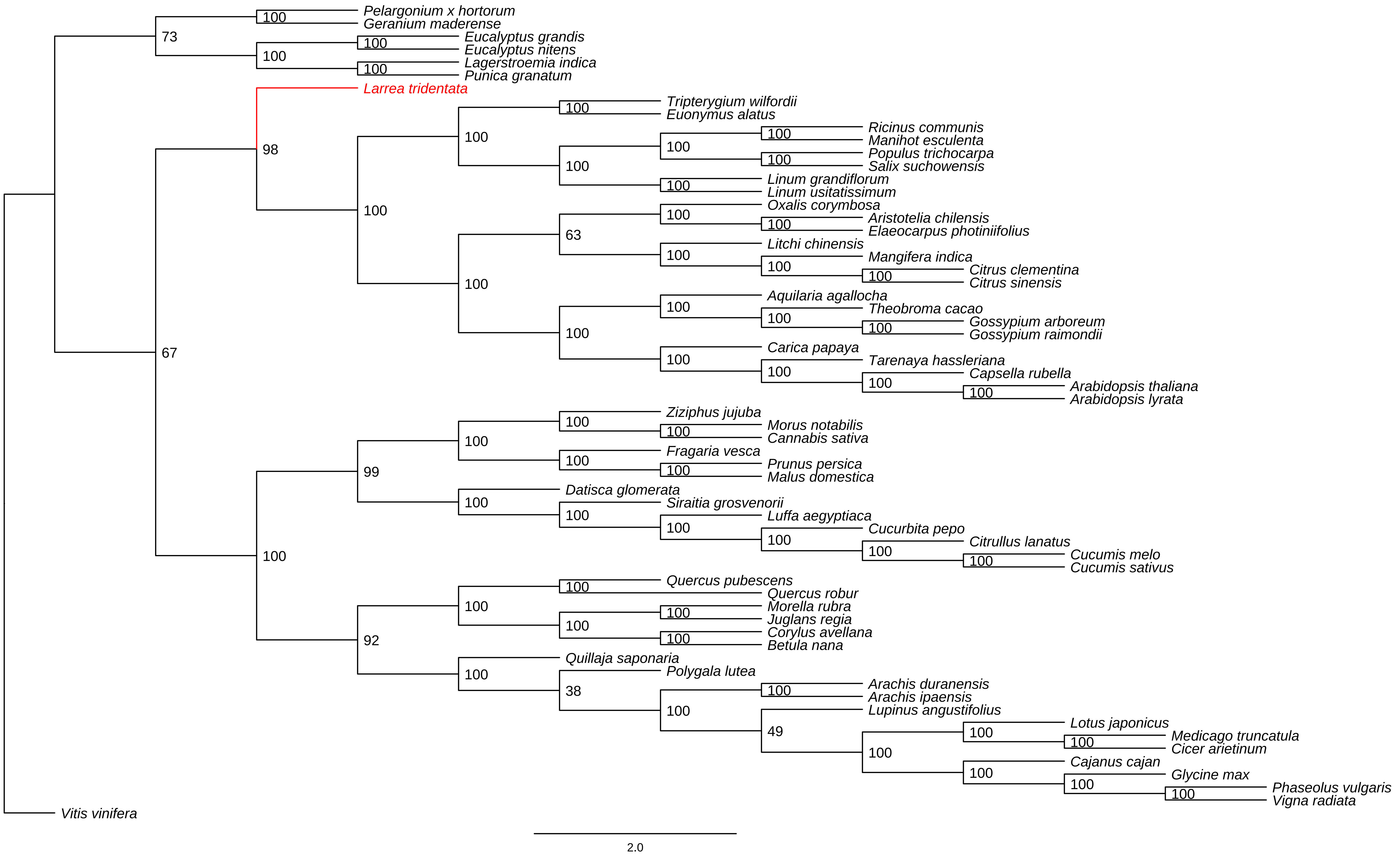

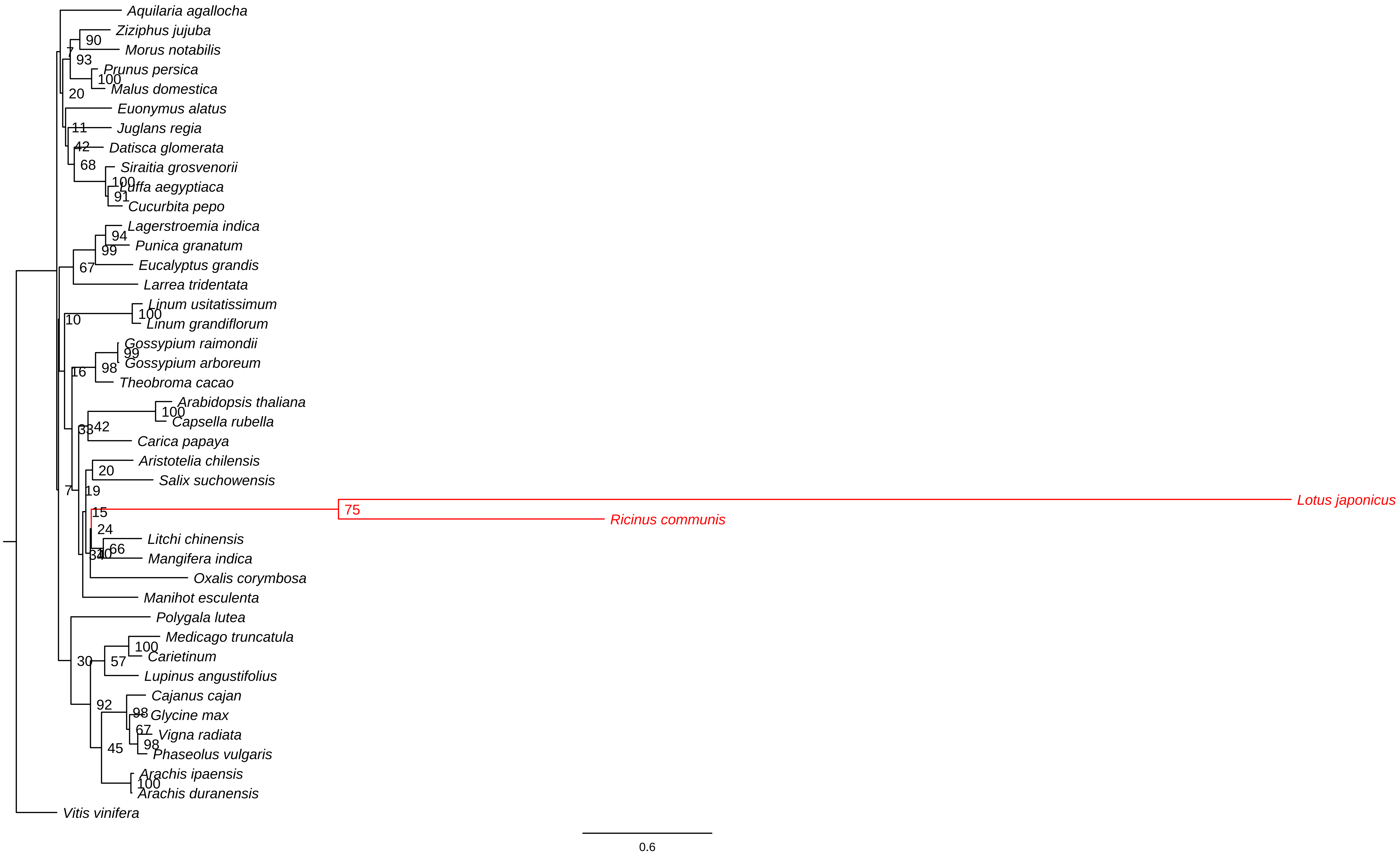

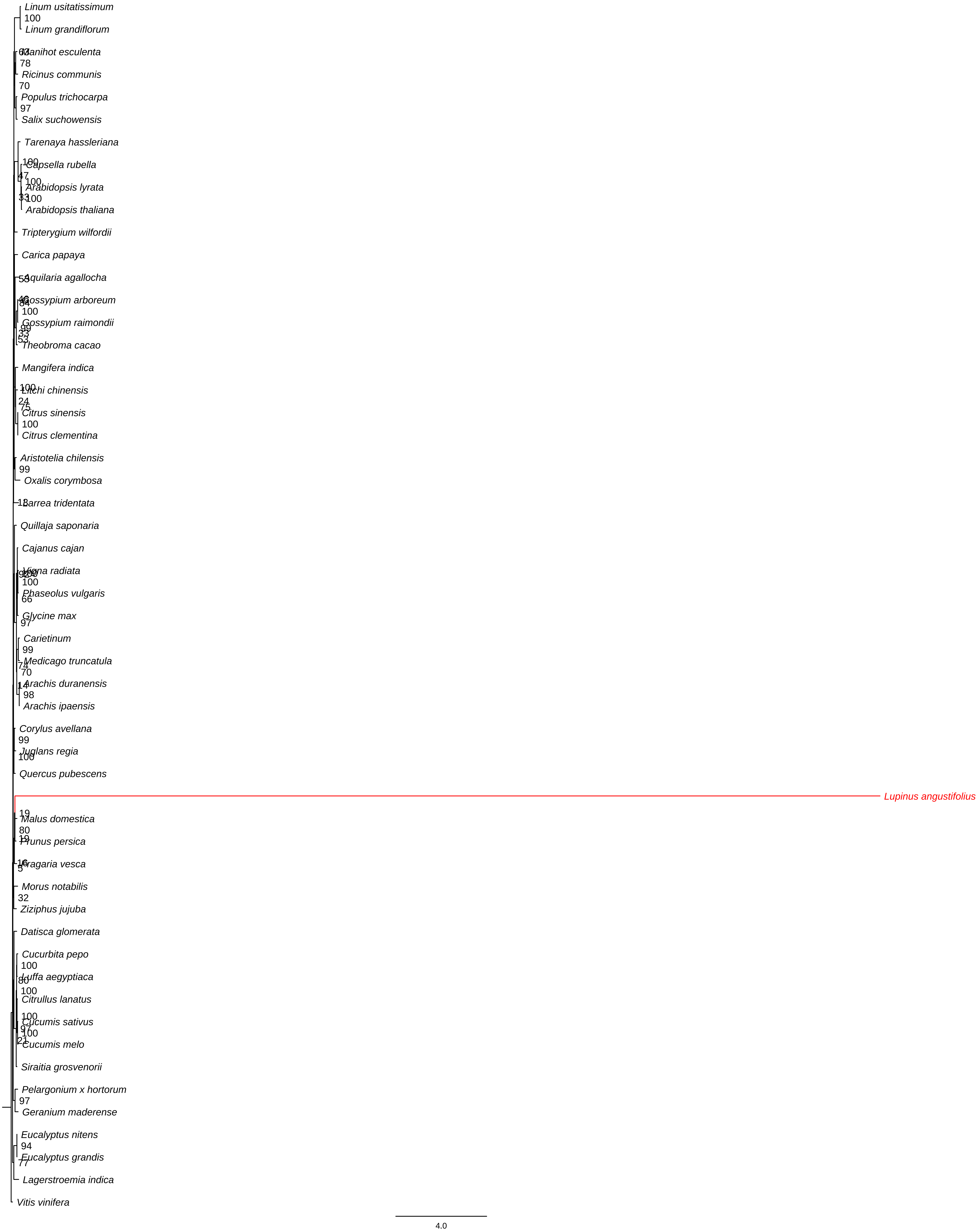

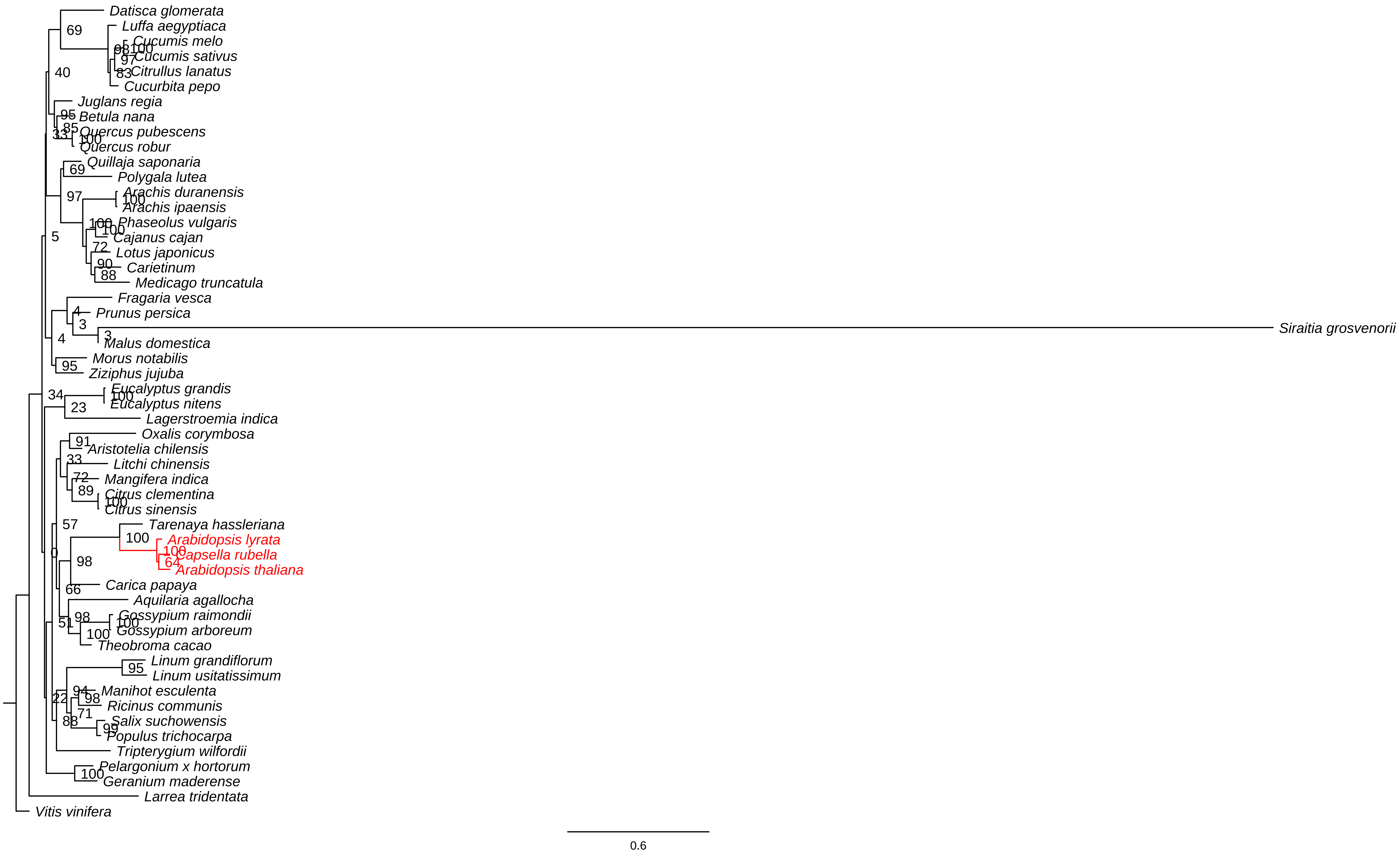